\documentclass[aps,onecolumn,preprintnumbers,amsmath,amssymb,nofootinbib,superscriptaddress,notitlepage]{revtex4-1}
\vspace{5mm}

\usepackage{slashed}
\usepackage{graphicx}
\usepackage{amssymb}
\usepackage{mathtools}
\usepackage{bbold}
\usepackage{amssymb,latexsym}
\usepackage{amsmath,amsbsy,bbm}
\usepackage{multirow}
\usepackage[vcentermath]{youngtab}
\usepackage{nicefrac}
\usepackage[perpage]{footmisc}
\usepackage{wrapfig,lipsum,booktabs}
\usepackage{caption}
\usepackage{subcaption}
\usepackage{graphicx}
\usepackage{cjhebrew}
\usepackage{cleveref}
\usepackage[utf8]{inputenc}
\usepackage{soul}

\usepackage{floatrow}
\usepackage[dvipsnames]{xcolor} 
\newfloatcommand{capbtabbox}{table}[][0.45\textwidth]

\newcommand{\eg}{\textit{e.g.}}
\newcommand{\ie}{\textit{i.e.}}

\newcommand{\ve}[1]{\ensuremath{\boldsymbol{#1}}}

\newcommand{\be}{\begin{equation}}
\newcommand{\ee}{\end{equation}}

\newcommand{\la}{\label}

\newcommand{\figref}[1]{fig.~(\ref{#1})}

\newcommand{\ccite}[1]{\cite{#1}}

\usepackage[normalem]{ulem}

\begin{document}

\title{Unitary interaction geometries in few-body systems}

\author{Lorenzo Contessi}
\address{
  IRFU, CEA, Université Paris-Saclay, 91191 Gif-sur-Yvette, France
}
\author{Johannes Kirscher}
\address{Department of Physics, SRM University - AP, Amaravati 522502, Andhra Pradesh, India}
\address{Theoretical Physics Division, School of Physics and Astronomy,
  The University of Manchester, Manchester, M13 9PL, UK}
\address{Institute for Nuclear Studies, Department of Physics,
The George Washington University, Washington DC 20052, USA}
  
\author{Manuel Pavon Valderrama}
\address{School of Physics, Beihang University, Beijing 100191, China} 

\date{\today}

\begin{abstract}
  We consider few-body systems in which only a certain subset
  of the particle-particle interactions is resonant.
  We characterize each subset by a {\it unitary graph}
  in which the vertices represent distinguishable particles
  and the edges resonant 2-body interactions.
  Few-body systems whose unitary graph is connected
  will collapse unless a repulsive 3-body interaction is included.
  We find two categories of graphs, distinguished by the kind of 3-body
  repulsion necessary to stabilize the associated system.
  Each category is characterized by whether the graph contains a loop or not:
  for tree-like graphs (graphs containing a loop) the 3-body force renormalizing
  them is the same as in the 3-body system with two (three)
  resonant interactions.
  We show numerically that this conjecture is correct for the 4-body case
  as well as for a few 5-body configurations.
  We explain this result in the 4-body sector qualitatively by imposing
  Bethe-Peierls boundary conditions on the pertinent
  Faddeev-Yakubovsky~decomposition of
  the wave function.
\end{abstract}

\maketitle

\section{Introduction}

Systems in which the 2-body scattering length is considerably larger
than the range of the interaction are called unitary or resonant.
They show properties that are independent of the details of their
interparticle interaction (provided it has a finite
range)~\ccite{Braaten:2004rn}, which is why
it is referred to as {\it universality}.
The reason is the presence of an exceedingly large separation of scales:
the ratio between the scattering length and any other length scale of
the system basically tends to infinity.
As a consequence, unitary 2-body systems can be described by a
parameter-free (or universal) theory.

The consequences of universality beyond the 2-body case
are more interesting and counterintuitive.
3-boson systems for which the 2-boson interaction is resonant exhibit
a characteristic geometric spectrum in which the ratio of the binding
energies of the $n$-th and $(n+1)$-th excited states
is approximately $(22.7)^2$.
This spectrum -- usually referred to as the Efimov effect -- was predicted
in the seventies~\cite{Efimov:1970zz} and confirmed experimentally
a decade and a half ago~\cite{Kraemer:2006}.
Similar geometric spectra have been predicted for larger
boson clusters~\cite{Carlson:2017txq}, systems of non-identical
particles~\cite{Helfrich:2010yr,PhysRevLett.108.073201,PhysRevLett.113.213201},
mass-imbalanced $P$- and even $D$-wave 3-body
states~\cite{Helfrich:2011ut}, etc.
From the point of view of symmetry, what is happening here
is that the continuous scale invariance of universal 2-body systems
becomes anomalous and breaks in the 3-body case as a consequence of
the quantization process, yet it survives as discrete scale invariance.
The unitary 3-body system is thus no longer parameter-free.
It acquires a 3-body parameter that can be identified with
the binding energy of the fundamental 3-body bound state.

This raises the question of what happens with systems of four or more particles
in the unitary limit.
We know~\cite{Bazak:2018qnu}~that the 4-body
parameter is not needed to predict the ground state of
the 4-bosons system and it only appears as a perturbative correction
together with finite range corrections.
However, this is not necessarily the case for all 4-body systems
if not all particles interact resonantly.
For 4-body systems of the $AABB$ type, with $A$ and $B$ denoting
two different species of particles (either bosons or distinguishable)
and where only the $AB$ interaction
is resonant, the 3-body parameters that are required to define the $AAB$ and $ABB$ subsystems
are insufficient to determine the binding energy of
the ground state~\cite{Contessi:2021ydv}.
Expressed differently, the $AABB$ system is a rare example
in which a 4-body parameter is required.

Conversely, not all universal few-body systems acquire a 3-body parameter.
The $P$-wave 3-body system with equal mass particles (or, equivalently,
the $AAB$ system if $A$ are fermions and $m_A = m_B$) does not
collapse or exhibit the Efimov effect.
For specific mass imbalances, this system forms 3-body bound states whose
binding energies depend only on the 2-body scattering
length~\cite{Kartavtsev_2007} and
only when the mass imbalance is large enough will it require
a 3-body parameter and display a geometric spectrum.
Another example is the $AABB$ system when the two species are fermions,
in which case there will be no bound state~\cite{PhysRevLett.93.090404}.

Going back to the non-fermionic case (here we consider distinguishable
particles), the present manuscript generalizes the methods and
findings of Ref.~\cite{Contessi:2021ydv} as follows:
instead of considering different types of 2-species clusters,
we will focus on the geometry of their resonant interactions.
In particular, we will characterize few-body systems in terms of a
{\it unitary graph}, here defined as a graph whose vertices and lines
represent, respectively, the particles and resonant
interactions of the system.
Provided the graph is connected, the few-body system requires
the definition of a 3-body parameter.
If the graph is a tree, the required 3-body parameter will be that of
the 3-body system with two unitary pairs (e.g. the $AAB$ or $ABB$
systems we discussed in the previous paragraph).
If the graph contains a cycle (a loop), the 3-body parameter of
the unitary 3-boson system will be needed instead.
We have tested the previous two statements explicitly
in the 4- and 5-body systems.

Furthermore, we expect that all the few-body system represented by
connected graphs will display the Efimov effect.
We conjecture that the geometric ratios between the binding
energies of the $n$-th and $(n+1)$-th bound state will be $(1986.1)^2$ and
$(22.7)^2$ for tree-like graphs and graphs containing cycles
respectively.
These ratios represent the ones that are found in the 3-body systems
when there are two and three unitary pairs~\cite{Naidon:2016dpf}.
The key assumption for this conjecture is that all systems with the same
Efimov ratios are renormalized by the same 3-body force.
We provide a heuristic argument explaining the scaling behavior
to be expected in certain $N>3$ systems as well as the type of
3-body force required to renormalize
the corresponding unitary graphs.

\section{Theory and methodology}
\label{sec:theory}

\subsection{Description of the $N$-body system}
\la{sec.theo.a}

We consider non-relativistic $N$-body systems described by
the Schr\"odinger equation
\begin{eqnarray}
\left(
-\frac{\hbar^2}{2m}\sum_i^N\vec{\partial}_i
+ \hat{v}
\right)~\Psi_{N}
 = 
E_N~\Psi_{N}
\;\; ,
\qquad
\la{eq.schroedinger}
\end{eqnarray}
where $\Psi_{N} = \Psi_{N} (\vec{r}_1, \ldots, \vec{r}_N)$ is
the wave function, $\vec{\partial}_i = \frac{{d}}{d \vec{r}_i}$
the derivative with respect to the coordinate of particle $i$,
$\hat{v}$ is the potential and $E_N$ the center-of-mass
energy of the $N$-body system.
We limit the discussion to distinguishable, equal-mass $m$ particles,
and hence no permutation symmetry is enforced on the wave function
$\psi_{1,\ldots,N}$.

For a general $N$-body system, the interaction potential $\hat{v}$ 
may include contain up to $N$-body forces.
  But for the unitary systems under scrutiny in the present manuscript
  it suffices to include 2- and 3-body forces, as higher order body forces
are not required for their description.
That is, we have:
\begin{eqnarray}
  \hat{v} = \sum_{i < j} v_{ij} + \sum_{i < j < k} w_{ijk} \, , 
\end{eqnarray}
with $v_{ij}$ and $w_{ijk}$ the 2- and 3-body potentials

Unitary 2-body systems are insensitive to the range of their interaction,
which is never probed.
As a consequence, the 2-body potential is effectively reduced
to a contact-range potential, \ie~a Dirac Delta in r-space.
This type of potential is singular and has to be regularized, \eg,
by including a cutoff in the calculations.
For concreteness we choose a Gaussian regulator in coordinate space:
\be
v_{ij} = v(\ve{r}_i,\ve{r}_j) 
= 
c_{ij}(\lambda) \,
e^{-\lambda^2\,\frac{(\ve{r}_i-\ve{r}_j)^2}{4}}
\;\;
,
\qquad
\la{eq.2pot}
\ee
where $\lambda$ is the cutoff (\ie, the auxiliary range we introduce
to make numerical calculations easier) and $c_{ij}$ a coupling constant.
This coupling depends on the cutoff (that is, $c_{ij} = c_{ij}(\lambda)$)
in such a way as to keep the 2-body system at unitarity
for arbitrary values of the cutoff $\lambda$,
provided the interaction of the particle pair $ij$ is unitary
in the first place.
The coupling and cutoff dependence is identical for every unitary 2-body
subsystem, and thus we write
\begin{eqnarray}
  c_{ij}(\lambda) = c(\lambda)\,f_{ij} \, , \label{eq:cij-to-c}
\end{eqnarray}
with $f_{ij} = 0$ or $1$ for non-unitary and unitary $ij$, respectively.
The cutoff dependence of regularized contact interactions
is well known~\cite{vanKolck:1998bw} and becomes
$c(\lambda) \propto \lambda^2$ in our specific case
(\ie, in our normalization of the regularized
Dirac Delta, see Eq.~(\ref{eq:pot-delta-reg})
for a more detailed explanation).

In practice we will calibrate $c(\lambda)$ by setting the $S$-wave scattering
length $a_0$ to exceed $10^5$ for each value of $\lambda$ considered.
The description of the system is independent of the cutoff,
as expected in a renormalized theory.

Unitary 3-body systems are, however, not insensitive to the range of
the interaction.
The reason is that the continuous scale invariance
of the 2-body system becomes anomalous in the 3-body system and
reduces to a discrete scale invariance.
To be explicit, while the 2-body system is invariant with respect to
$\vec{r} \to \kappa \vec{r}$ transformations for arbitrary $\kappa$,
in the 3-body system, this symmetry only survives for $\kappa = \kappa_0$
with $\kappa_0$ a specific real number (\eg, the famous
$\kappa_0 \approx 22.7$ scaling factor for the Efimov effect
in the 3-boson system~\cite{Efimov:1970zz}).

In practical terms, this manifests as a cutoff dependence of the ground state
of the 3-body system, whose binding energy will diverge as $\lambda^2$
(the reason is that $\lambda$ is the only momentum scale in the system).
The inclusion of a 3-body force stabilizes the energy of the 3-body
ground state and removes the unphysical cutoff
dependence~\cite{Bedaque:1998kg,Bedaque:1999ve}.
With a Gaussian regulator the 3-body force reads
\begin{eqnarray}
w_{ijk} = w(\ve{r}_i,\ve{r}_j,\ve{r}_k) = 
d_{ijk}(\lambda)\,  
e^{-\lambda^2 \left[ \frac{(\ve{r}_i-\ve{r}_j)^2}{4} + \frac{(\ve{r}_i-\ve{r}_k)^2}{4}\right]}
\, , \nonumber \\
\la{eq.3pot}
\end{eqnarray}
where $d_{ijk}(\lambda)$ and its running are determined by the condition of
reproducing the ground state (or an arbitrary excited state) of the $ijk$
3-body system (this requires that at least two pairs of particles within
the $ijk$ set are unitary).
If we assume that the ground-state energy of every bound
3-body subsystem is the same, we can make the additional
simplification
\begin{eqnarray}
  d_{ijk}(\lambda) = d(\lambda)\,g_{ijk} \, , \label{eq:dijk-to-d}
\end{eqnarray}
with $g_{ijk} = 0$ or $1$ depending on the particular $ijk$ 3-body
subsystem under consideration (we will specify
this in the following lines).

\subsection{Characterization of the few-body configurations as unitary graphs}
\la{sec.theo.c}

We are interested in few-body systems where not all of the $N$ particles
interact resonantly, but only a subset of them.
Without loss of generality, the rest of the pairs (\ie, the non-unitary ones)
are considered to be non-interacting: in principle their interaction
can be treated as a perturbative correction around the unitary limit
set by the unitary pairs.
The reason is that the scattering lengths of the non-unitary pairs are
arbitrarily smaller than the scattering lengths of
the unitary pairs.
However, owing to the breakdown of continuous scale invariance for $N \geq 3$,
for the previous expansion to be valid there is the additional proviso
that the ratio between the 3-body scale (\eg, the characteristic
length scale or size of the 3-body bound state) and any scale
associated with the residual non-resonant interactions
(\eg, the scattering length of the non-unitary
pairs of particles) should be large.

Here, we describe (partially) unitary $N$-body systems in terms of a
{\it unitary graph}, a graph in which vertices correspond to
particles and lines to unitary interactions.
The sets of 3- and 4-body configurations we consider and
their names are shown in~\figref{tab.3topos}~and
~\figref{tab.4topos}.
These are all the possible connected graphs
with three and four vertices.
For the graphs with three vertices (\figref{tab.3topos}) we denote them as
\begin{enumerate}
\item[(3a)] {\it delta} or $\Delta$ (for its resemblance to
  the Greek letter), which is also the fully connected graph
  with three nodes and
\item[(3b)] {\it lambda} or $\Lambda$ (again, relating to the Greek letter).
\end{enumerate}
For 4-vertex graphs (\figref{tab.4topos}) we introduce
\begin{enumerate}
\item[(4a)] the {\it full} (the fully connected graph with four nodes),
\item[(4b)] the {\it circle-slash}\footnote{Of all shown graphs, this is the only one not amenable to a
straight-forward $N$-vertex generalization.} or simply {\it slash} (which is also referred
  to as the {\it diamond} in the literature~\cite{graphbook}),
\item[(4c)] the {\it circle}, 
\item[(4d)] the {\it line} (self-explanatory),
\item[(4f)] the {\it paw}, as it is often referred to
  (other names are the {\it 3-pan} graph or
  the {\it (3,1)-tadpole} graph\footnote{A $k$-pan is a graph with $(k+1)$ points, where $k$ of the points are connected in such a
  way that they form a cycle and with the odd point connected to one of the points within that $k$-cycle (see fig.~(1.2) in Ref.~\cite{graphbook});
  a $(k,m)$-tadpole
  is a $k$-cycle where one of the points within the cycle is connected to a tail of $m$ connected
    points (\ie, an $m$-line).}), and
\item[(4g)] the {\it star}, which we previously named {\it dandelion}
  in~\cite{Contessi:2021ydv}, and also referred to as
  the {\it claw}~\cite{graphbook}.
\end{enumerate}
A few of the previous graphs are easily generalizable to the $N$-body case.
But, if not stated otherwise, we will be referring to the 4-body
version of the graph.
In some cases, we might indicate the $N$-body generalizations
by adding the number of points in the graph to its name, \eg,
the 5-full, the 6-circle, the 7-line, or the 8-star graphs.

The $N$-body potential $\hat{v}$ associated with these systems/graphs is
fully determined by the set of $ij$ pairs and $ijk$ triplets for which
the 2- and 3-body interaction is non-zero,
\ie, the $ij$ and $ijk$ for which the $f_{ij}$ and $g_{ijk}$ factors defined
in Eqs.~(\ref{eq:cij-to-c}) and (\ref{eq:dijk-to-d}) are equal to one,
$f_{ij} = 1$ and $g_{ijk} = 1$.
For the 2-body forces this is trivial: $f_{ij} = 1$ if the pair $ij$ are
connected by a line  when we look at the corresponding unitary graph (and
$f_{ij} = 0$ if the pair is not connected).

For the 3-body forces, the characterization of the $ijk$ triplets for
which $g_{ijk} = 1$ is relatively simple for the $\Delta$ and $\Lambda$
three body systems: $g_{123} = 1$.
However, it becomes more involved in the 4-body case,
at least in principle (in practice, concrete calculations accept
  a very convenient simplification that we will comment later).
The explicit definition of the $g_{ijk}$ coefficients depends
on whether or not a particular graph contains a $\Delta$ subgraph:
\begin{itemize}
\item[(i)] if it contains a $\Delta$ subgraph, $g_{ijk} = 1$ 
  if and only if $f_{ij} = 1$, $f_{jk} = 1$ and $f_{ik} = 1$, \ie,
  if the particle pairs $ij$, $jk$ and $ik$ are all unitary
\item[(ii)] while if there is no $\Delta$ subgraph, it will be only
  necessary that two of the $ij$, $jk$ and $ik$ pairs are unitary.
\end{itemize}
That is, there is a 3-body force for each $\Delta$ triplet,
  or if there is none, for each $\Lambda$ triplet.
Alternatively, we might define the two following sets of
relevant 2- and 3-body interactions for the previous graphs:
\begin{eqnarray}
  \mathcal{F}_{\rm graph} &=&
  \left\lbrace(i,j):i,j\in\lbrace 1,\ldots,N\rbrace,~
  i<j,f_{ij} \neq 0 \right\rbrace \, ,\qquad\la{eq.set} \, \\
  \mathcal{G}_{\rm graph} &=&
  \Big\lbrace(i,j,k):i,j,k\in\lbrace 1,\ldots,N\rbrace, ~i<j<k, \nonumber \\
  && \quad g_{ijk} \neq 0 \Big\rbrace \, ,
\end{eqnarray}
With $\mathcal{F}_{\rm graph}$ and $\mathcal{G}_{\rm graph}$
it is possible to redefine the potential as
\be
\hat{v}=\sum_{(i,j)\in\mathcal{F}_{\rm graph}} v_{ij} +
\sum_{(i,j,k) \in \mathcal{G}_{\rm graph}} w_{ijk} \, ,
\qquad
\la{eq.potcfg}
\ee
where the definitions of $v_{ij}$ and $w_{ijk}$ do not involve dependence on
the indices for the couplings:
\begin{eqnarray}
  v_{ij} &=& c(\lambda)\,e^{-\lambda^2 \frac{(\ve{r}_i-\ve{r}_j)^2}{4}} \, , \\
  w_{ijk} &=& d(\lambda)\,
  e^{-\lambda^2 \left[ \frac{(\ve{r}_i-\ve{r}_j)^2}{4} + \frac{(\ve{r}_i-\ve{r}_k)^2}{4}\right]} \, .
\end{eqnarray}
That is, with the definition of $\mathcal{F}_{\rm graph}$ and
$\mathcal{G}_{\rm graph}$ we can now dispense of the $f_{ij}$ and
$g_{ijk}$ factors previously found
in Eqs.~(\ref{eq:cij-to-c}) and (\ref{eq:dijk-to-d}).

However, numerically we observe that the inclusion of the
3-body force on all the triplets, even the not strictly necessary ones to avoid collapse, does not affect the predictions for the 4-body ground-state energies.
We can explain this by the fact that contact range three-body forces have little impact on non-collapsing triplets of particles.
Thus, for practical reasons, all numerical simulations consider the 3-body
interaction acting on all triplets (\ie, $g_{ijk} = 1$ for all $i, j, k$).
For instance,  the 4-line results were obtained with a 3-body interaction
between the two edges and one of the interior nodes.

\subsection{Practical implementation}
\la{sec.theo.b}

For all calculations in this work, we set $\hbar=c=1$,
and the particle mass to unity: $m=1$.
The parameter $\lambda$ and all energies are given in mass units $[m]$.
The coupling strength $c(\lambda)$ is chosen to induce an $S$-wave scattering
length of $1 / a_0 = 10^{-5}\,m$ for $\lambda<100~m$.
We calibrated the 3-body coupling $d(\lambda)$ as to generate a 3-body
ground state with binding energy $E_3 = -0.01~m$.
This numerical value for $E_3$ was chosen for both 3-body configurations
of \figref{tab.3topos}, that is, for the $\Lambda$ (two resonant pairs)
and $\Delta$ (three resonant pairs) configurations.

All conclusions in this work are drawn from the dependence of the ground-state
solution of Eq.~(\ref{eq.schroedinger}) on the cutoff $\lambda$ and
on the unitary graph representing the $N$-body system.
We calibrate $c(\lambda)$ by solving the 2-body Schr\"odinger equation at
zero energy with a standard Numerov integration algorithm.
For the 3-body coupling $d(\lambda)$ and for the calculations of the ground-state
energies $E_N$ for $N \geq 3$, we use two different variational methods,
both of which optimize a set of Gaussian basis functions:
the stochastic-variational method (SVM)~\cite{Suzuki:1998bn}
and the refined-resonating-group method (RGM)~\cite{Hofmann:1986}.
All $N\geq3$ calculations employed the SVM, and we used the RGM
as an additional verification of the values of $E_N$
in the 3- and 4-body systems for a subset of cutoffs.   

\subsection{Cutoff dependence of the couplings}
\la{sec.theo.d}

\begin{figure}
\includegraphics[width=\textwidth]{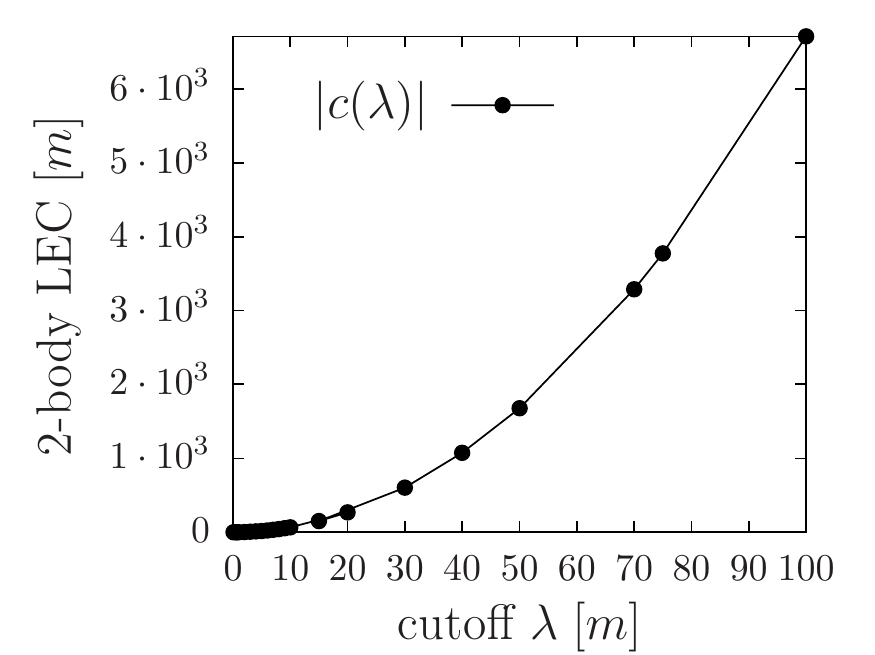} 
\caption{Cutoff $\lambda$ dependence of the coupling strength $c$
of a Gaussian-regulated contact interaction between a pair of particles
with an $S$-wave scattering length $a_0\approx10^5\,m^{-1}$.}
\label{fig.LECc}
\end{figure}

With the parameters and methods specified, we obtain the expected, above-mentioned
cutoff dependence
for the 2-body strength:
$c(\lambda) \propto \lambda^2$ (see~\figref{fig.LECc}).
In our notation, the $c$ absorbs factors stemming from the
Gaussian regularization of the Dirac Delta, \ie,
\begin{eqnarray}\label{eq:pot-delta-reg}
  v_{ij} &=& C(\lambda)\,\delta_{\lambda}(\vec{r}_i - \vec{r}_j) \nonumber \\
  &=& C(\lambda)\,\frac{\lambda^3\,e^{-\lambda^2 \frac{(\vec{r}_i - \vec{r}_j)^2}{4}}}
  {8\pi^{3/2}} \, ,
\end{eqnarray}
from which the relation between $c(\lambda)$ and the more commonly used
$C(\lambda)$
\begin{eqnarray}
  c(\lambda) = \frac{\lambda^3}{8\pi^{3/2}}\,C(\lambda) \, , 
\end{eqnarray}
is obtained.
If we take into account that
$C(\lambda)=-\frac{2\pi^2}{m \lambda}\theta^{-1}$~\cite{Konig:2016utl},
with $\theta$ a regulator-dependent number, it is apparent that
we should indeed have $c(\lambda) \propto \lambda^2$.

\begin{figure}
\begin{tabular}{cc}
 lambda($\Lambda$) & delta($\Delta$)\\ 
 \includegraphics[width=55mm]{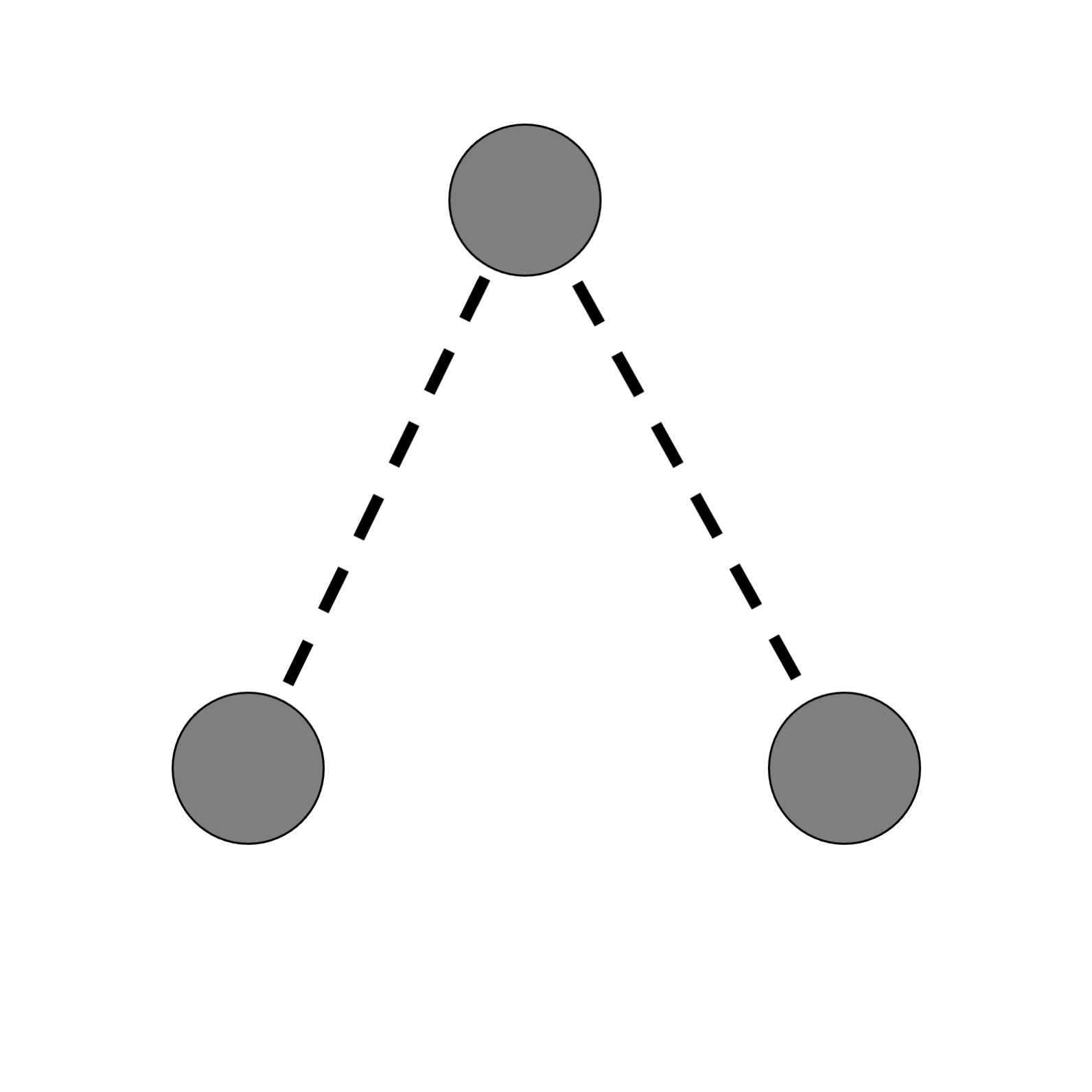} &
 \includegraphics[width=55mm]{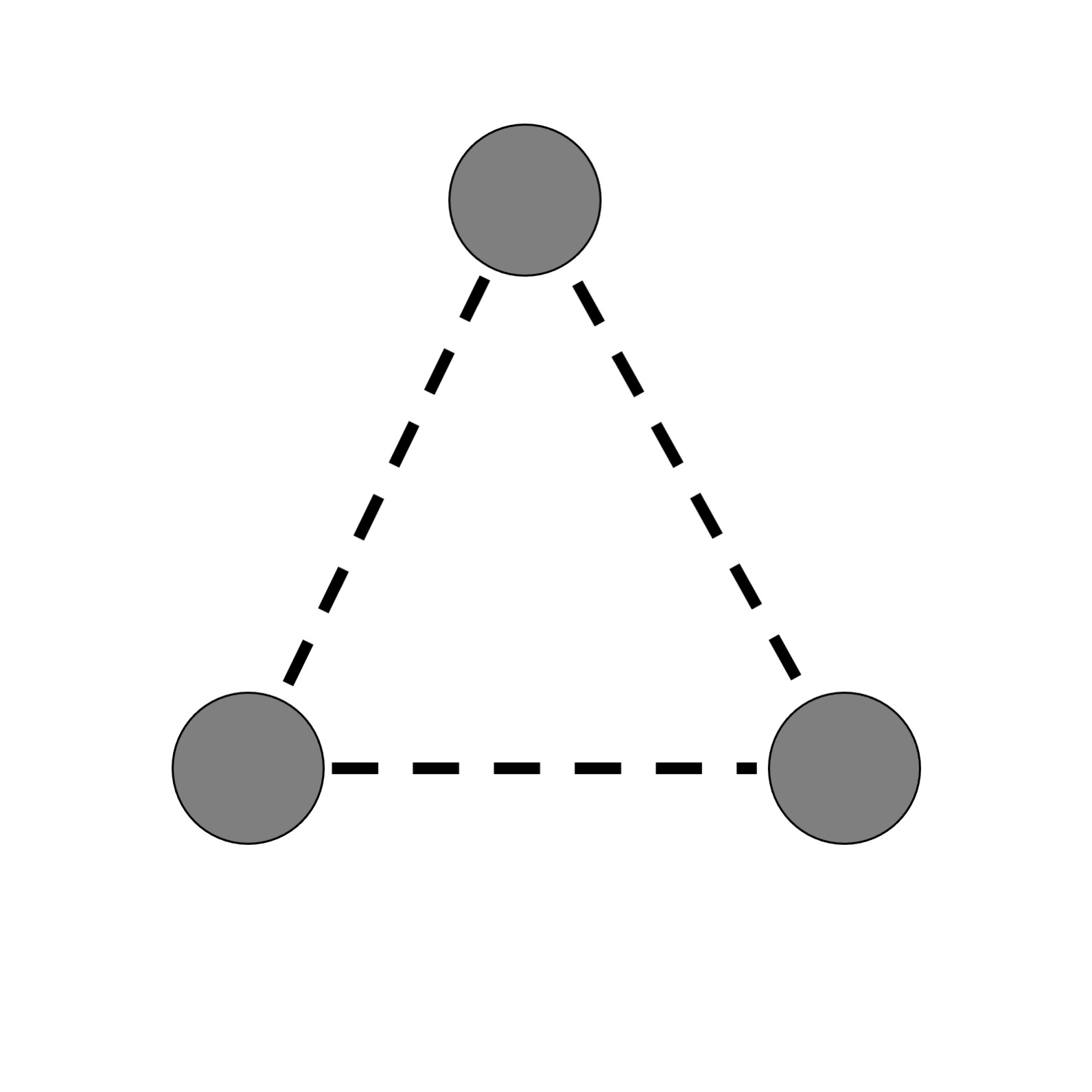} \\
\end{tabular}
\caption{3-body systems with 2 ($\Lambda$) and 3 ($\Delta$) resonant pairs:
  the gray circles and dashed lines represent the distinguishable particles
  and their unitary interactions, respectively.}
\label{tab.3topos}
\end{figure}

If we use the 2-body potential with the coupling of~\figref{fig.LECc}
and in the absence of a 3-body force,
the calculated 3-body ground-state energy diverges as the square of
the cutoff, \ie, $E_3 \propto \lambda^2$.
This is explicitly shown in~\figref{fig.3collapse}~for the two configurations
of~\figref{tab.3topos}~-- the $\Delta$ and $\Lambda$ configurations -- where
a parabolic fit to these numerical results finds that
\be
E_{3\Lambda}\simeq -0.000118\,\lambda^2
\quad\text{and}\quad E_{3\Delta} \simeq -0.0596\,\lambda^2 \, .
\qquad
\label{eq.3collapse}
\ee
That is, we find a more rapid collapse of the fully resonant $\Delta$
configuration compared with the $\Lambda$ configuration.

\begin{figure}
\begin{tabular}{c}
  \includegraphics[width=\textwidth]{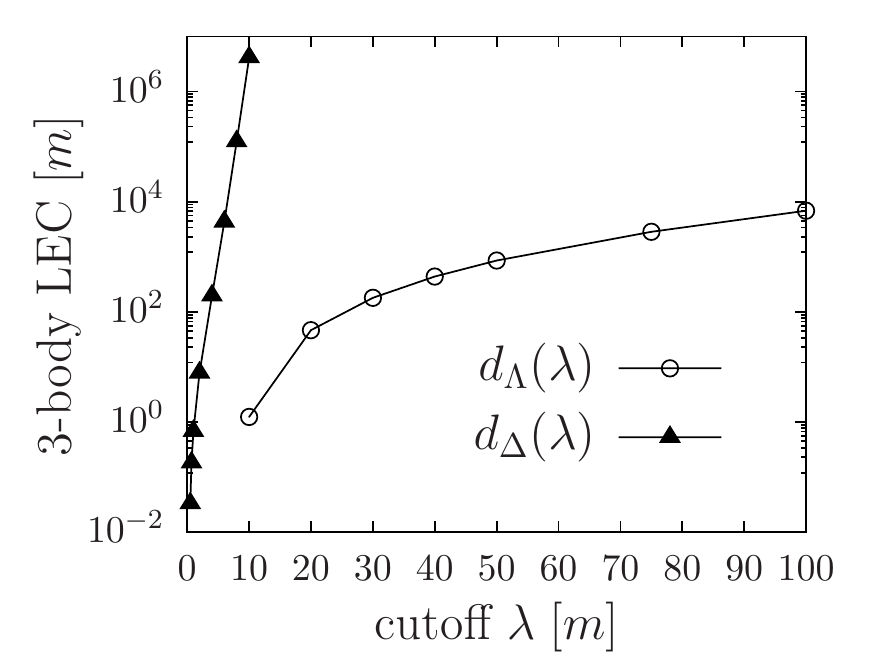} \\
\end{tabular}
\caption{The running of the 3-body coupling strength $d(\lambda)$
which renormalizes the fully resonant (filled triangles)
and the 2-pair resonant (empty circles) 3-body
systems to a single bound state with $E_3=0.01~m$.
}
\label{fig.LECd}
\end{figure}

These 3-body collapses are avoided by including the 3-body contact terms
in Eq.~(\ref{eq.3pot}).
The coupling $d(\lambda)$ is calibrated under the condition that the energy of
the 3-body ground state remains constant with the cutoff.
By choosing $E_{3\Delta} = E_{3\Lambda} = - 0.01~m$ for both the delta and
lambda configurations, we find
numerically\footnote{Our choice for the cutoff interval,
$\lambda<10~m$ for $d_{\Delta}(\lambda)$ and
$\lambda<100~m$ for $d_{\Lambda}(\lambda)$,
follows numerical constraints. The collapse of the unrenormalized system
is expressed in separately diverging ground-state expectation values of
the kinetic energy and 2-body potential operators.
The latter does so more rapidly with $\lambda$, and hence, the
diverging binding energy is itself a result of the cancellation of
two even larger numbers.
The repulsion required thus from a 3-body counterterm demands
strengths which approach numerical limits faster
for the fully resonant system (see~\figref{fig.LECd}).}
the $d_{\Delta}(\lambda)$ and $d_{\Lambda}(\lambda)$
behaviors of \figref{fig.LECd}.

Even though we limit ourselves to the calculation of the ground state of
the $\Delta$ and $\Lambda$ trimers, the excited states of these systems
do in principle display the Efimov effect~\cite{Efimov:1970zz}.
The discrete scale invariance that characterizes the Efimov effect manifests
differently for the $\Delta$ and $\Lambda$ 3-body
systems~\cite{Naidon:2016dpf}:
while for the delta configuration, we have the well-known scaling
$E^{(n+1)}_{3\Delta} / E^{(n)}_{3\Delta} \simeq (22.7)^2$,
for the lambda configuration, the scaling is
$E^{(n+1)}_{3\Lambda} / E^{(n)}_{3\Lambda} \simeq (1986.1)^2$ instead.
The superscript $(n)$ indicates the $n$-th
excited state of the trimer.
We have not checked these two
geometric factors that characterize the Efimov spectra specifically in this work.
Yet, the $\Delta$ configuration collapses more rapidly
than the $\Lambda$, which reflects the smaller discrete scaling factor of
the former ($22.7$ for the $\Delta$) compared to the later
($1986.1$ for the $\Lambda$).

\begin{figure}
\begin{tabular}{c}
 \includegraphics[width=\textwidth]{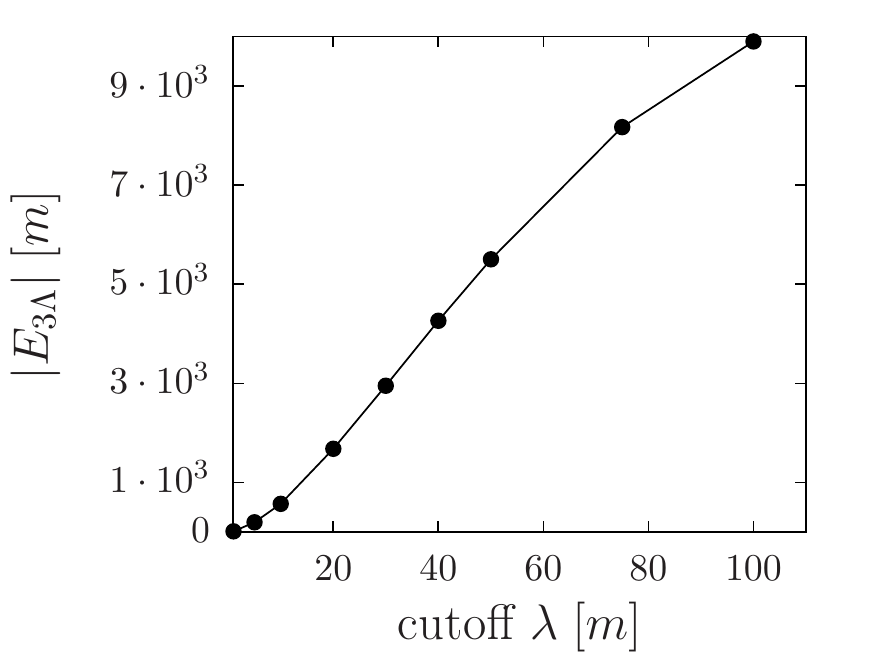} \\
\end{tabular}
\caption{
  Divergent behavior of ground-state energy $E_{3\Delta}$ of the $\Delta$ 3-body
  system (three resonant pairs) when we include the 3-body force that
  renormalized the $\Lambda$ 3-body system (two resonant pairs). 
}
\label{fig.3fullcollapse}
\end{figure}

\begin{figure}
\begin{tabular}{c}
  \includegraphics[width=\textwidth]{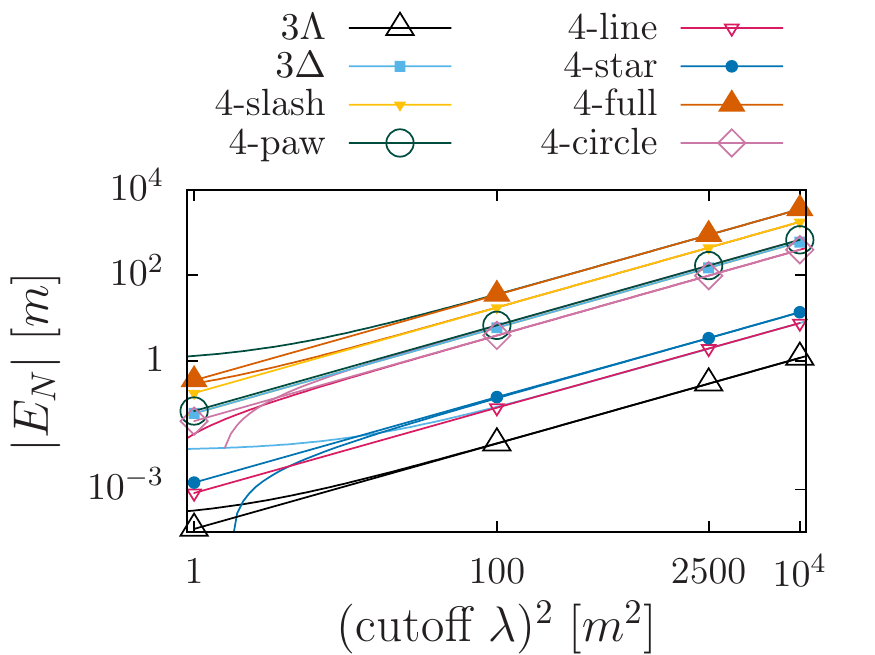}  \\
\end{tabular}
\caption{
  Regulator dependence of the 3- and 4-body ground-state
  binding energies without any collapse-preventing 3-body interaction.
  The lines highlight the $\propto\lambda^2$
  dependence of the binding energies of
  the states.
}
\label{fig.3collapse}
\end{figure}

\section{The 4-body sector}
\label{sec:4-body}

\subsection{Renormalization}

\begin{figure*}
\begin{tabular}{cc}
  full
  &
  slash
  \\
 \includegraphics[width=55mm]{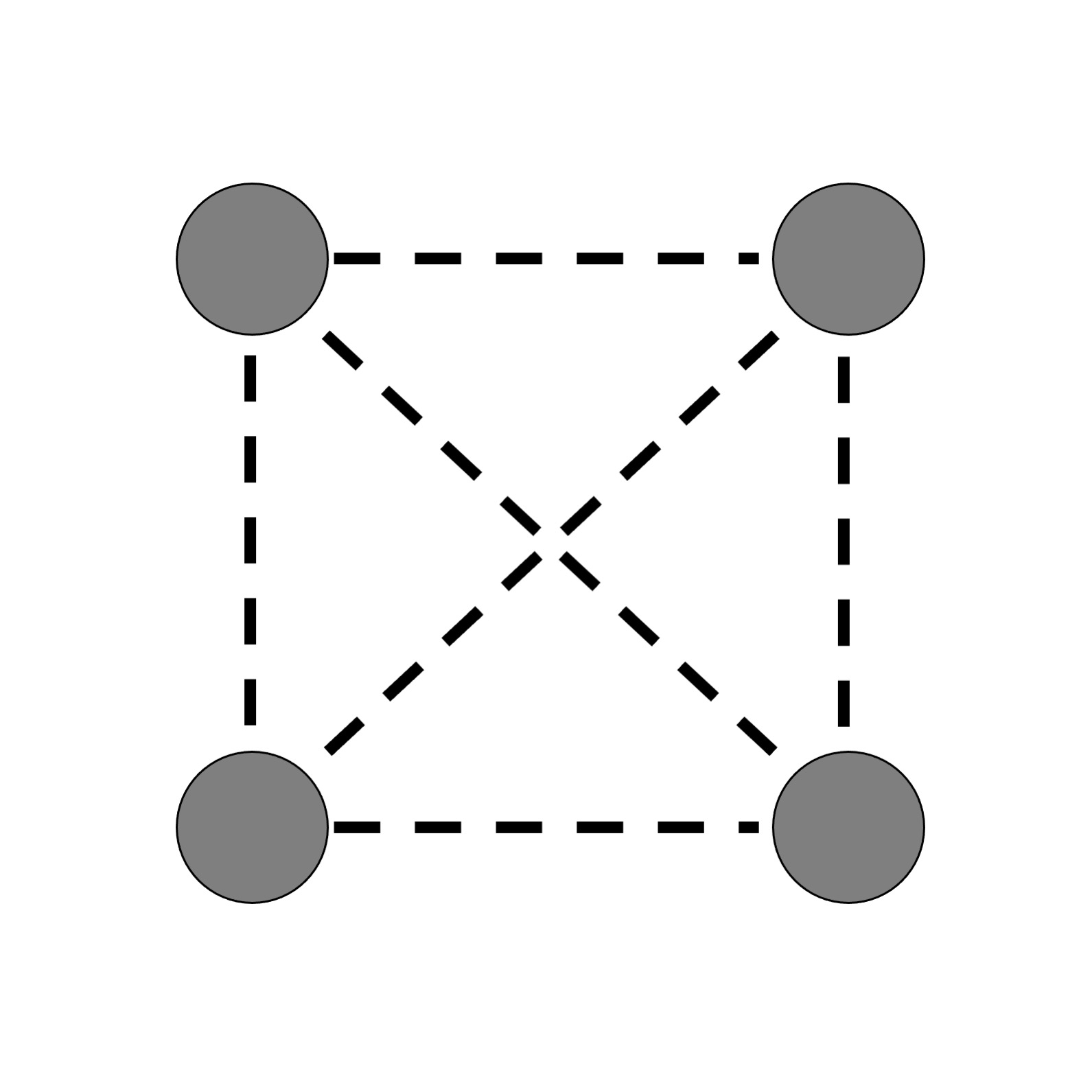} &   \includegraphics[width=55mm]{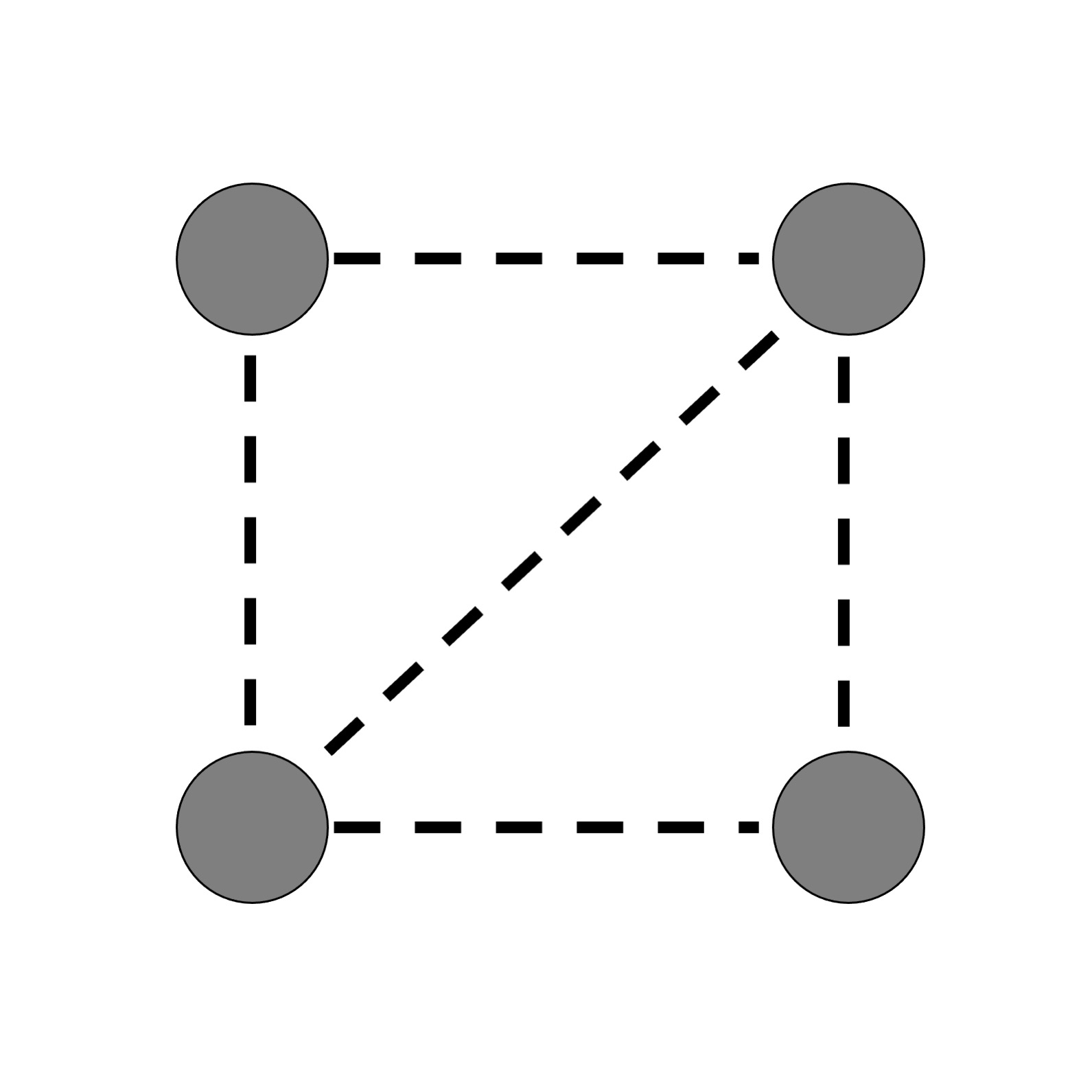} \\
 circle
 & 
 line
 \\
 \includegraphics[width=55mm]{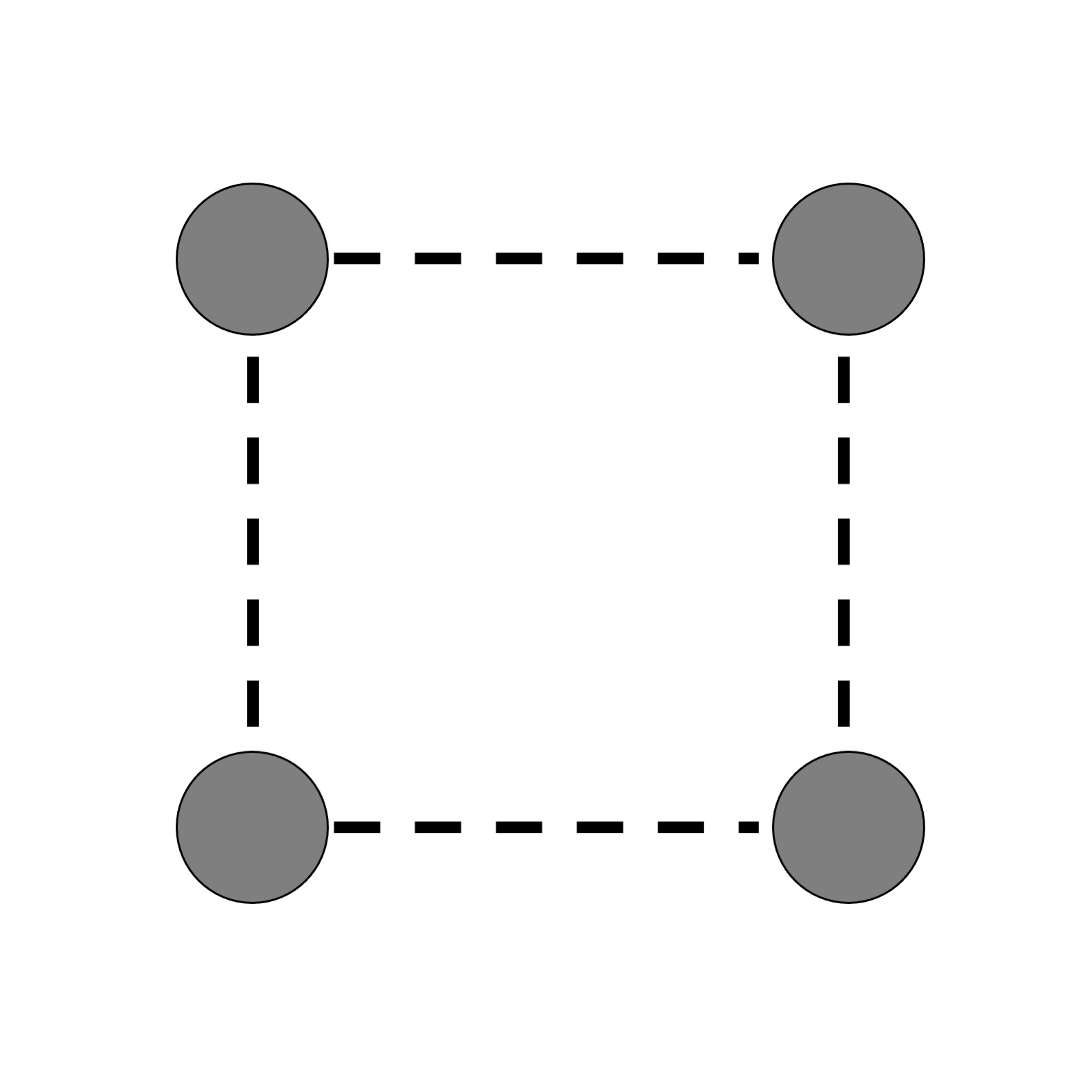} &   \includegraphics[width=55mm]{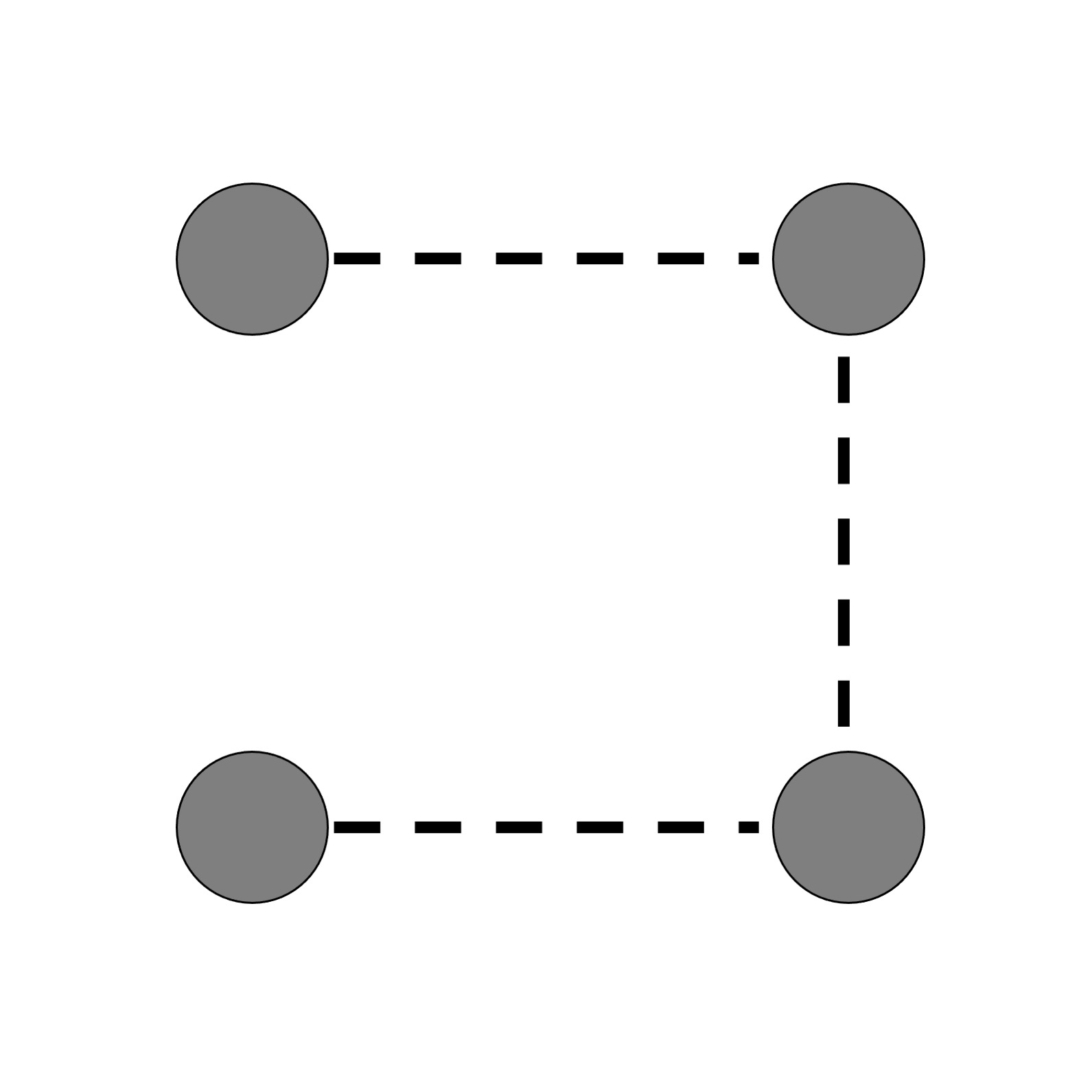} \\
 paw
 & 
 star
 \\
 \includegraphics[width=55mm]{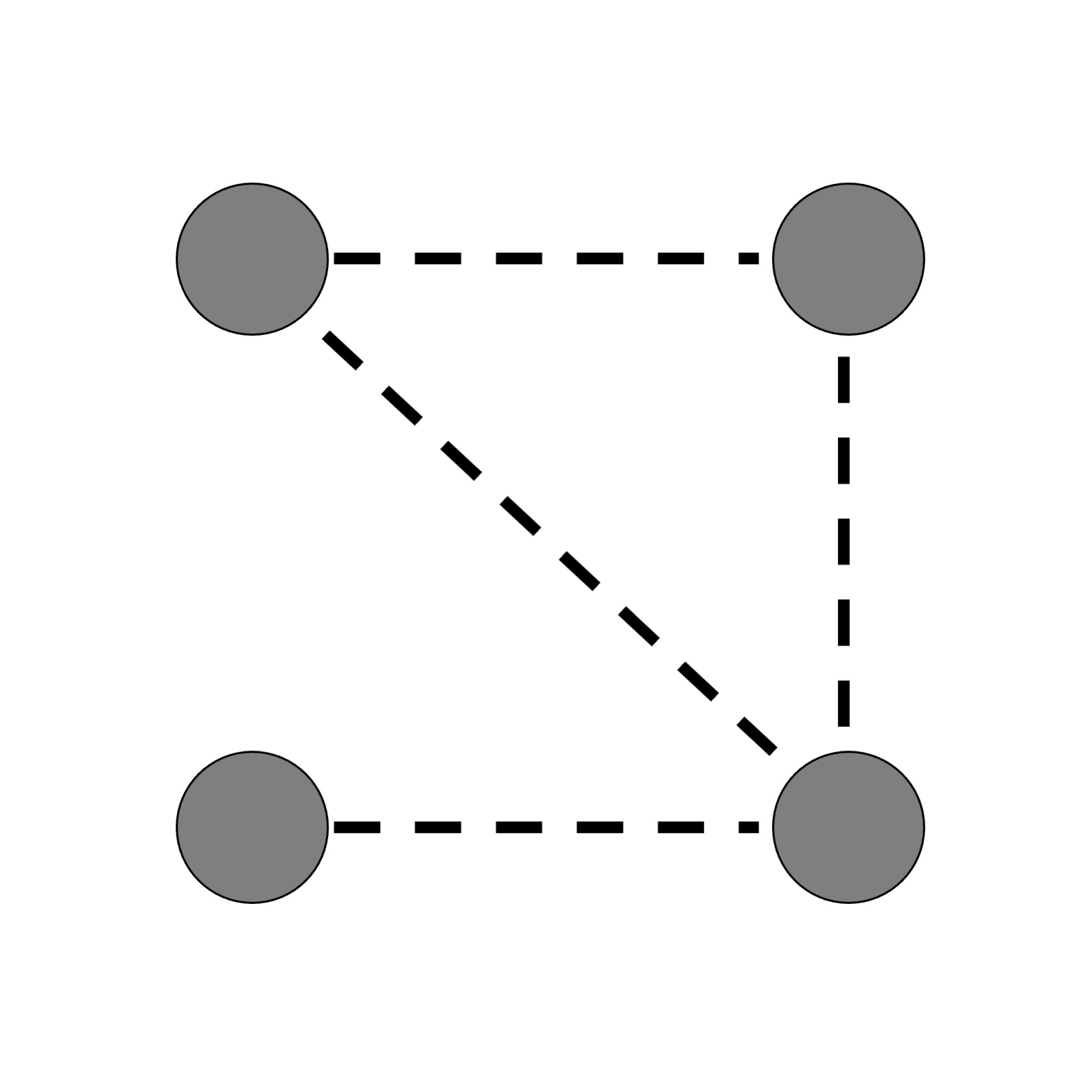} &   \includegraphics[width=55mm]{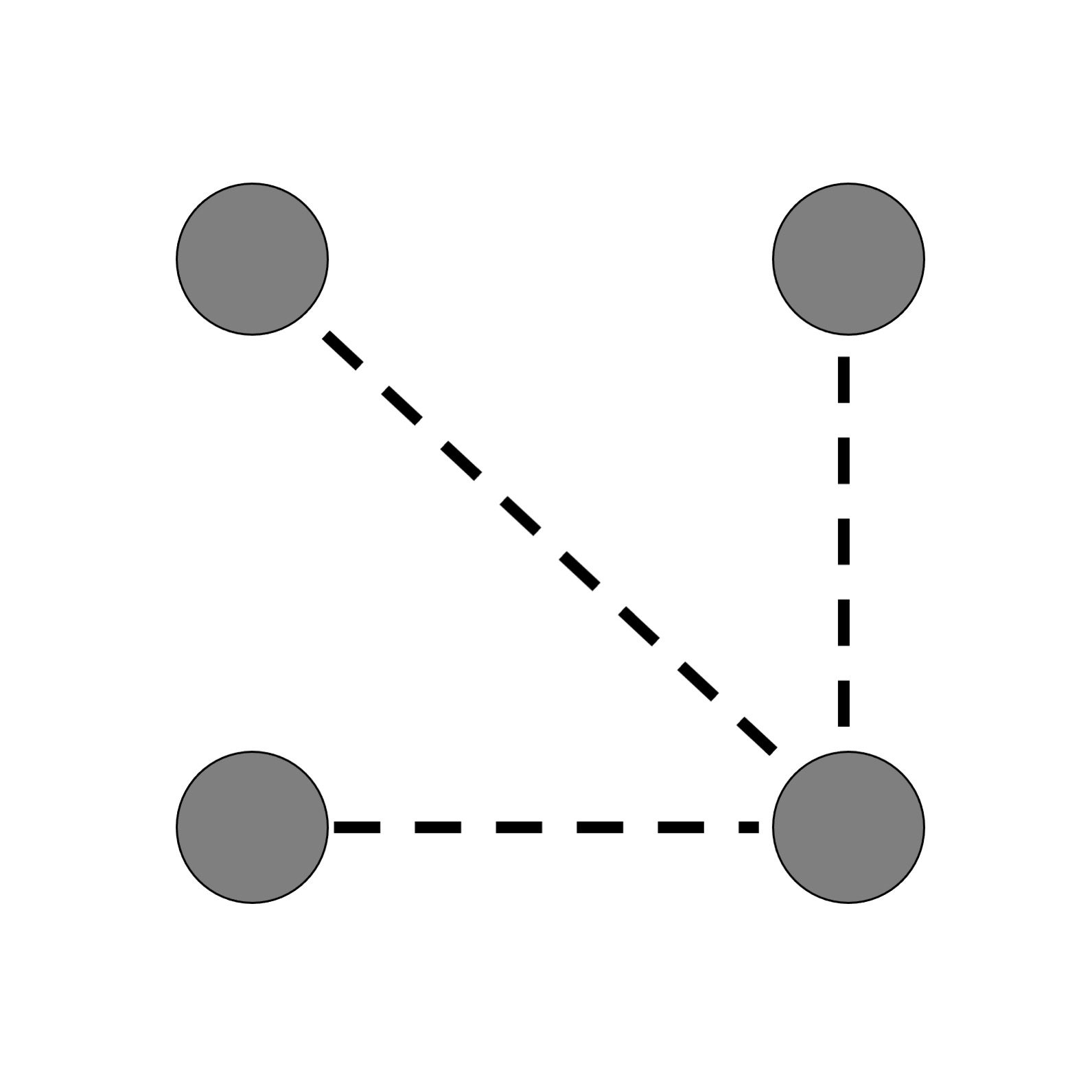} \\
\end{tabular}
\caption{Configurations of 4 distinguishable particles (gray vertices)
  for which their resonant pair interactions (dashed lines) form
  a connected graph.}
\label{tab.4topos}
\end{figure*}

Next, we consider all unitary 4-body systems in which the resonant
2-body interactions form a connected graph.
There are six such configurations which we list in~\figref{tab.4topos}.

In the absence of 3-body forces, the ground-state energy of
each of these 4-body configurations exhibits the quadratic collapse
with respect to the regulator $\lambda$ that
we encountered in the 3-body case, Eq.~(\ref{eq.3collapse}).
The reason is the presence of at least one $\Delta$ or $\Lambda$ subgraph
  in each of the connected 4-body graphs.
  These subgraphs will collapse in the absence of 3-body forces
  as discussed in the previous section.

\begin{figure}
\begin{tabular}{c}
  \includegraphics[width=\linewidth]{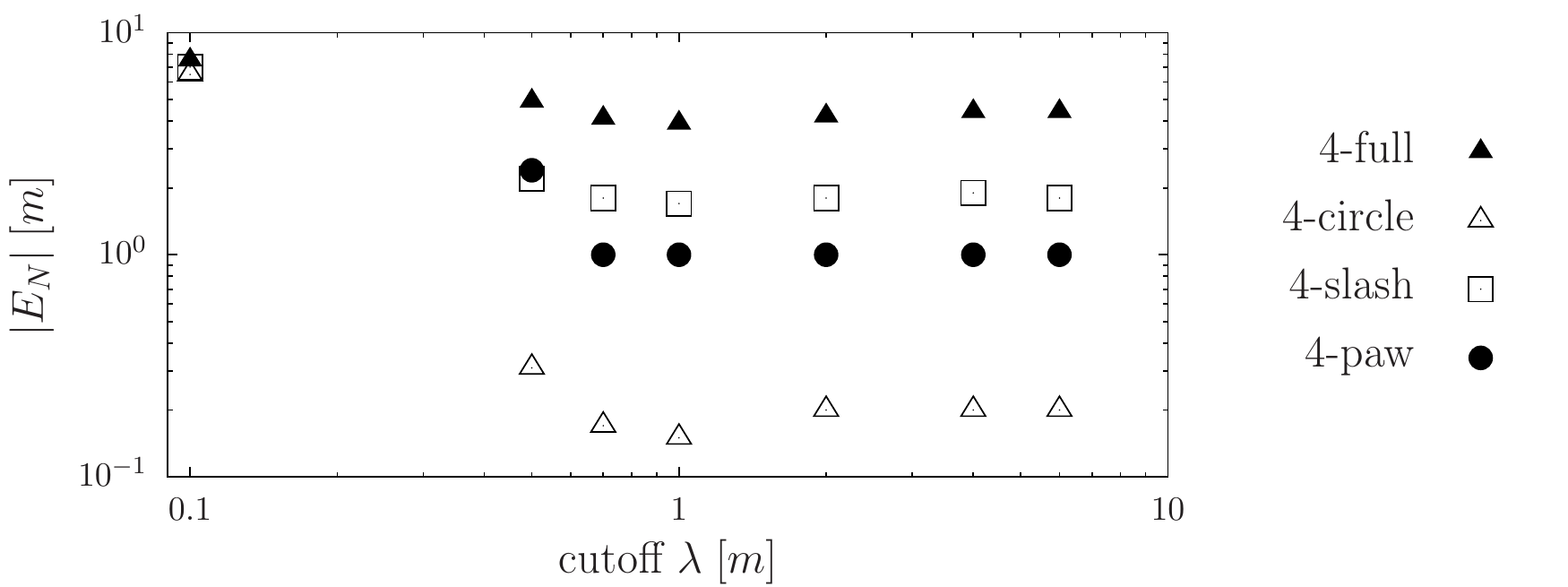} \\
  \includegraphics[width=\linewidth]{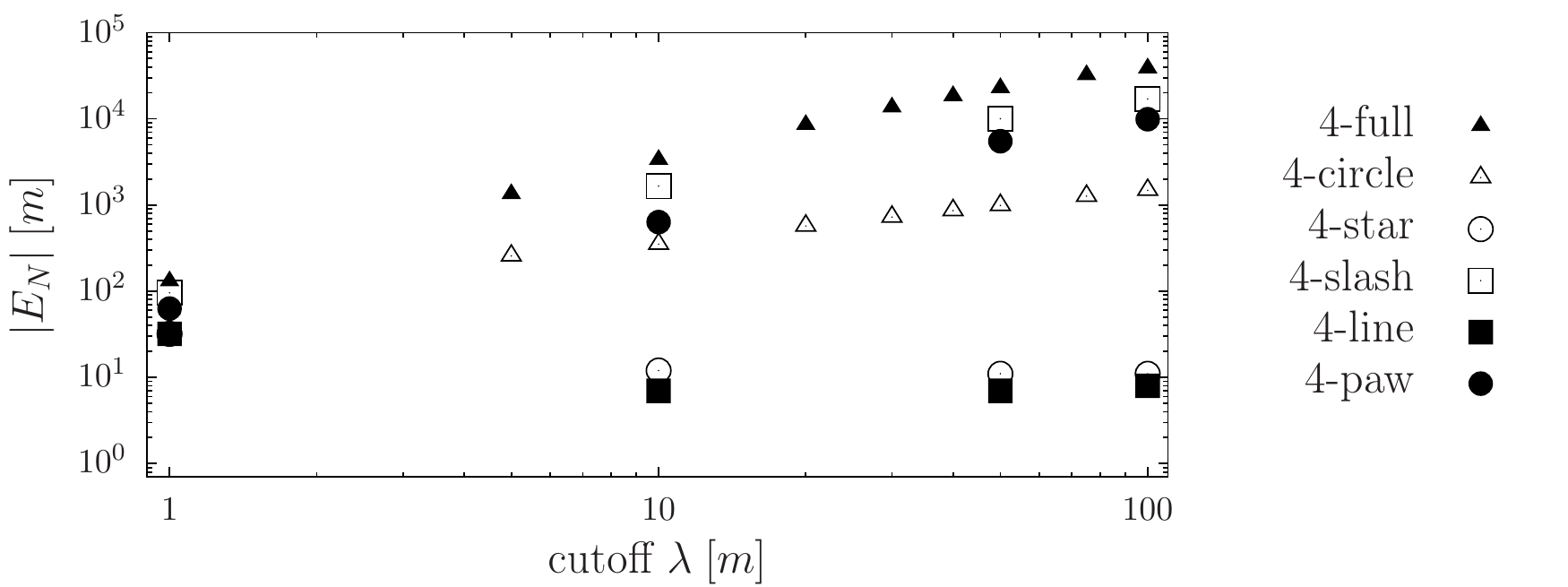}
\end{tabular}
\caption{Regulator dependence of the ground-state energy of the six 4-body configurations
if the running of the 3-body counter-term is set in the fully-resonant 3-body system
(upper panel, $d_\Delta$) and in the 2-pair-resonant 3-body system (lower panel, $d_\Lambda$).} 
\label{fig.4gs}
\end{figure}

The inclusion of a 3-body force (see Sec.~\ref{sec:theory})
solves the problematic collapse.
However, there is an ambiguity regarding which 3-body force to use:
as shown in the previous section, the 3-body coupling $d(\lambda$)
defined in Eq.~(\ref{eq.3pot}) has two possible solutions depending
on whether the 3-body systems is $\Delta$- or $\Lambda$-shaped.
In our calculations, only one choice for the running of $d$ with $\lambda$
was able to properly renormalize the system and generate
a finite binding energy.
Our numerical results for employing the two runnings in the various
graphs (\figref{tab.4topos}) are summarized graphically in~\figref{fig.4gs}.
From these results, we infer 
the following correspondence between a unitary graph
and the running of the 3-body force:
\begin{itemize}
\item[(i)] the full, slash, paw and circle graphs require $d_{\Delta}(\lambda)$,
\item[(ii)] the line and the star graphs require $d_{\Lambda}(\lambda)$.
\end{itemize}
This is a rather intuitive result, except for the circle:
naively, we would expect that the renormalization of a 4-body system
will only require a $\Delta$-type 3-body coupling $d_{\Delta}(\lambda)$
if its unitary graph contains one or more $\Delta$-shaped subgraphs.
Conversely, if there are only $\Lambda$-type subgraphs,
a $d_{\Lambda}(\lambda)$ coupling should suffice.
This is indeed what happens in five of the six configurations where,
as we will see, this behavior can be understood in terms of
a relatively simple heuristic argument grounded on
the Bethe-Peierls boundary conditions for each of
these systems.
The circle is a remarkable exception: numerically we find that it
requires $d_{\Delta}(\lambda)$ -- the 3-body coupling that renormalizes
the $\Delta$-shaped 3-body system -- although it does not contain
$\Delta$ subgraphs.

In the upper panel of~\figref{fig.4gs} we show the cutoff dependence
of the 4-body binding energy for the configurations
that require $d_{\Delta}(\lambda)$.
The quantitative ratios we find are:
\begin{eqnarray}
E_{{\rm 4-full}} &=& 4.4(1) \, E_{3\Delta} \, , \\
E_{{\rm 4-slash}} &=& 1.8(1) \, E_{3\Delta} \, ,\\
E_{{\rm 4-paw}} &\approx& 1.0 \, E_{3\Delta} \, , \la{eq.4to3ratiosc}  \\
E_{{\rm 4-circle}} &\approx& 0.2 \, E_{3\Delta} \, ,
\qquad
\end{eqnarray}
where for the full configuration we reproduce the well-known relation
between the binding energy of the unitary 3- and 4-boson system.
For the circle, the 4-body binding energy is smaller than the 3-body
$\Delta$-type configuration.
This is not a problem because the circle cannot decay into a 3-body
$\Delta$ bound system and a free particle.
Notice that the line and star configurations are not shown in the upper panel
of~\figref{fig.4gs} simply because when $d_{\Delta}(\lambda)$ is
applied to them it happens to be too repulsive to allow
for a bound state below the 3-body threshold.

Now, if we use the $d_{\Lambda}(\lambda)$ coupling instead, only
the line and star configurations converge, while the full, slash, paw,
and circle collapse (see lower panel of~\figref{fig.4gs}).
For the line and star configurations, we find
\begin{eqnarray}
E_{{\rm 4-star}} &=& 11(1) \, E_{3\Lambda} \, , \la{eq.4to3ratios2a} \\
E_{{\rm 4-line}} &=& 8(1) \, E_{3\Lambda} \, . \la{eq.4to3ratios2b}
\qquad
\la{eq.4to3ratios}
\end{eqnarray}
These ratios are not of order one. They are also certainly larger
than the respective ratios for the full, slash, paw, and circle.

\subsection{Combinatorial approximation to binding}

The renormalized binding energy ratios that we have calculated
reveal an intriguing pattern: they are proportional to
the number of interacting triplets
in that system.
First, we consider the $\Delta$-like configurations (except the 4-circle,
which is the 4-body system that behaves in a more peculiar way).
The number of interacting triplets is
\begin{eqnarray}
  4 \,:\, 2 \,:\, 1 \, ,
\end{eqnarray}
for the full, slash, and paw configurations, respectively.
This is to be compared with the ratios of their binding energies:
\begin{eqnarray}
  4.4(1) \,:\, 1.8(1) \,:\, 1 \, .
\end{eqnarray}
For the $\Lambda$-like configurations (star and line)
the number of interacting triplets is
\begin{eqnarray}
  3 \,:\, 2 \, ,
\end{eqnarray}
while the ratios of their binding
energies are
\begin{eqnarray}
  2.8(6) \,:\, 2 \, ,
\end{eqnarray}
and thus in agreement with the ratios of interacting triplets.
The uncertainties result from propagating the errors in the binding energies
of the different configurations.

Another prediction that can be derived from the combinatorial
argument is noteworthy: the binding energy of the
$N$-body full configuration.
For this prediction, we have to first consider the paw configuration, \ie,
a $\Delta$ graph with a fourth particle
interacting resonantly with one of the particles forming
the $\Delta$.
Owing to the resonant nature of this interaction, the na\"ive expectation is that
the binding energy of the odd particle with the $\Delta$ 3-body subsystem
will be just below the $\Delta$ plus free particle threshold,
which is precisely what is expressed by Eq.~(\ref{eq.4to3ratiosc}).
In fact, this argument could be extended for an $N$-body system composed of
a $\Delta$ followed by a line of $N-3$ particles -- the $N$-paw or the
$(3,N-3)$-tadpole graph -- for which the binding energy will
approximately be that of the $\Delta$ 3-body system:
\begin{eqnarray}
  E_{N{\rm -paw}} \approx 1.0\,E_{3\Delta} \, .
\end{eqnarray}
It is possible that this relation becomes less accurate as the tail of the tadpole, \ie,
the chain of particles attached to the $\Delta$ subgraph, becomes longer.

If we return now to the full $N$-body system, the previous assumption
together with the combinatorial hypothesis leads to the following ansatz:
\begin{eqnarray}
  E_{N{\rm -full}} &\approx& \begin{pmatrix} N \\ 3 \end{pmatrix}\,E_{N{\rm -paw}}
  \approx \frac{N (N-1) (N-2)}{6}\,E_{3 \Delta} \nonumber \\
  &\approx& \{ 1, 4, 10, 20, 35, 56\, \dots \}\,E_{3 \Delta} \nonumber \\
  && \quad \mbox{for $N=3,4,5,6,7,8\,\dots$} \, , 
\end{eqnarray}
where in the second line we specified the values
for different numbers of particles.
This approximation eventually breaks as $N$ increases.
For contact-range forces, in general, we expect that for high enough $N$
the binding energy of these systems will display
{\it saturation}, that is, a binding energy proportional to
the number of particles $N$.
As the ground state of the {\it full} $N$-body system is as bound as
the unitary $N$-boson system~\footnote{The reason
  is the permutation symmetry of the full $N$-body system, which effectively
  implies that the $N$ particles behave as identical particles. Even though
  this allows both symmetric and antisymmetric configurations
  (bosonic and fermionic behaviors), the ground state will
  correspond to a fully symmetric configuration.},
we can compare our results with the ratios obtained for the latter,
which have been extensively studied in the literature.
If we use the ratios obtained in Ref.~\cite{vonStecher:2009qw}
\begin{eqnarray}
  E_{N{\rm -full}} &=& \{ 1.0, 4.7, 10.6, 18.6, 27.9, 38.9\, \dots \}\,E_{3 \Delta} \nonumber \\
  && \quad \mbox{for $N=3,4,5,6,7,8\,\dots$} \, , 
\end{eqnarray}
then we see that even though the combinatorial argument works well
for $N \leq 6$ eventually its uncertainties end up increasing 
with $N$ (about $18, 6, 7, 21, 31\%$ for $N=4,5,6,7,8$).
If we use instead the more recent calculation of~\cite{Bazak:2016wxm},
the ratios for $N = 3,4,5$ will be in better agreement
with our approximation
\begin{eqnarray}
  E_{N{\rm -full}} &=& \{ 1.0, 4.2, 9.5, 16.3\, \dots \}\,E_{3 \Delta} \nonumber \\
  && \quad \mbox{for $N=3,4,5,6\,\dots$} \, , 
\end{eqnarray}
though we are limited to $N \leq 6$ in this case.
Finally, as the number of particles grows, saturation properties emerge
(\ie, $E_{N{\rm -full}}$ becomes proportional to $N$,
as has been shown in~\cite{Carlson:2017txq}) and
our combinatorial approximation will cease to be valid.

\section{The 5-circle and the unitary loop conjecture}
\label{sec:5-body}

Next, we consider 5-body unitary systems.
In this case there are a large number of connected unitary graphs -- 21
configurations\footnote{The number of connected graphs
  for $N = 2,3,4,5,6,7,8,\dots$ is $1,2,6, 21, 112, 853, 11117, \dots$
  (integer sequence A001349 in~\cite{oeis};
  see~\cite{osti_4180737}~for an explicit calculation
  of their number up to $N = 18$).} -- and for practical reasons
we will focus on a few unitary geometries only.

\begin{figure}
\begin{tabular}{ccc}
 5-full & 5-star & 5-circle \\
 \includegraphics[width=55mm]{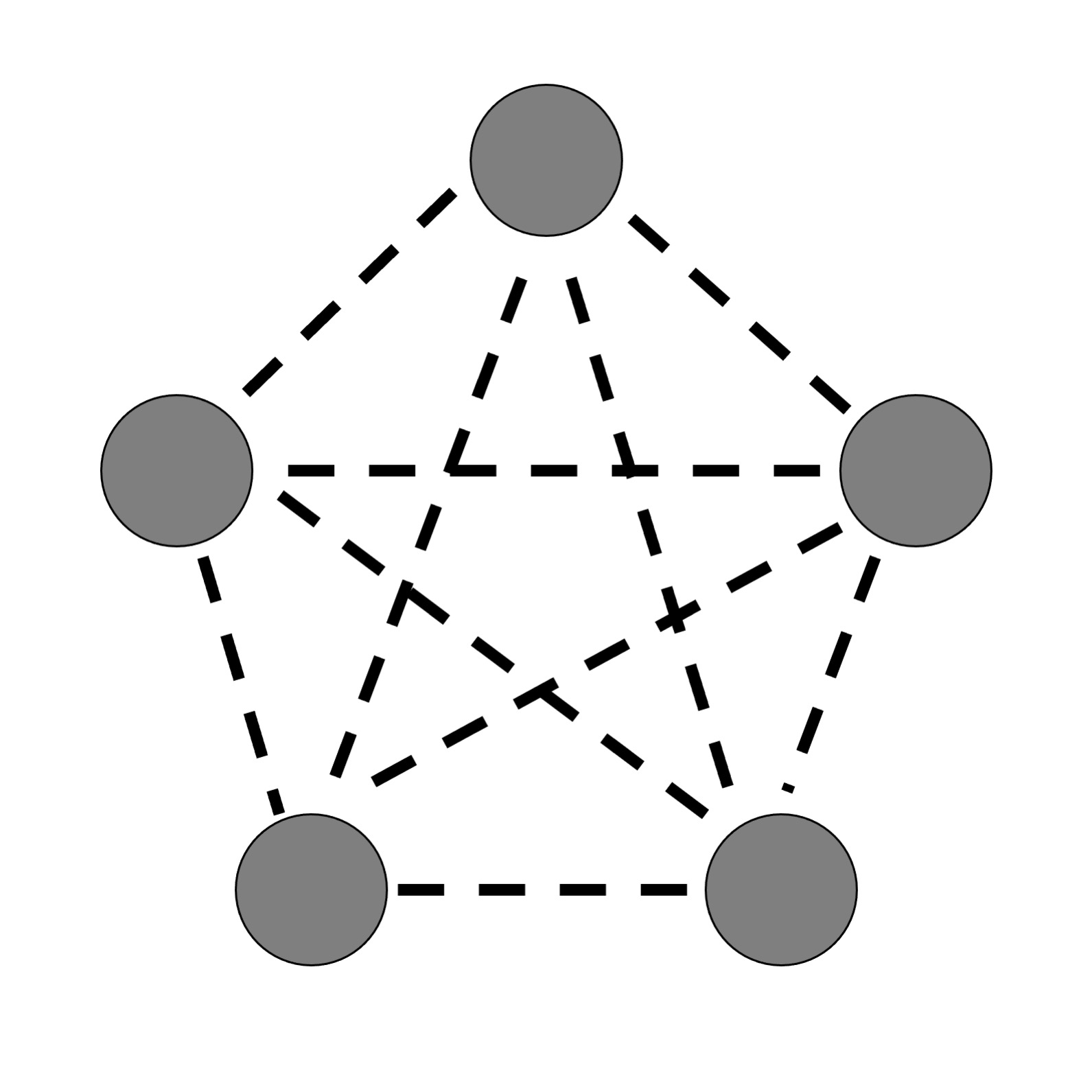} & 
 \includegraphics[width=55mm]{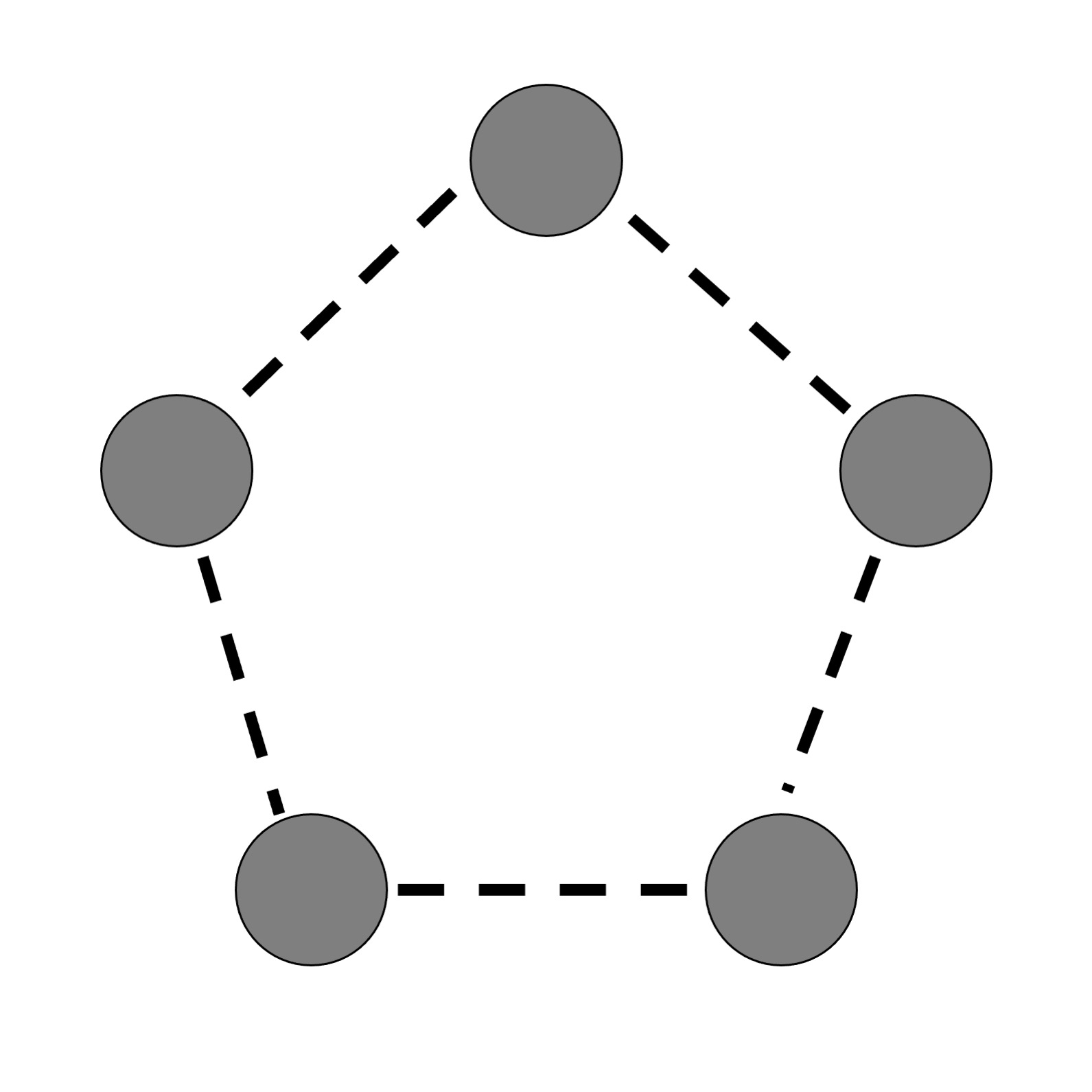} &
 \includegraphics[width=55mm]{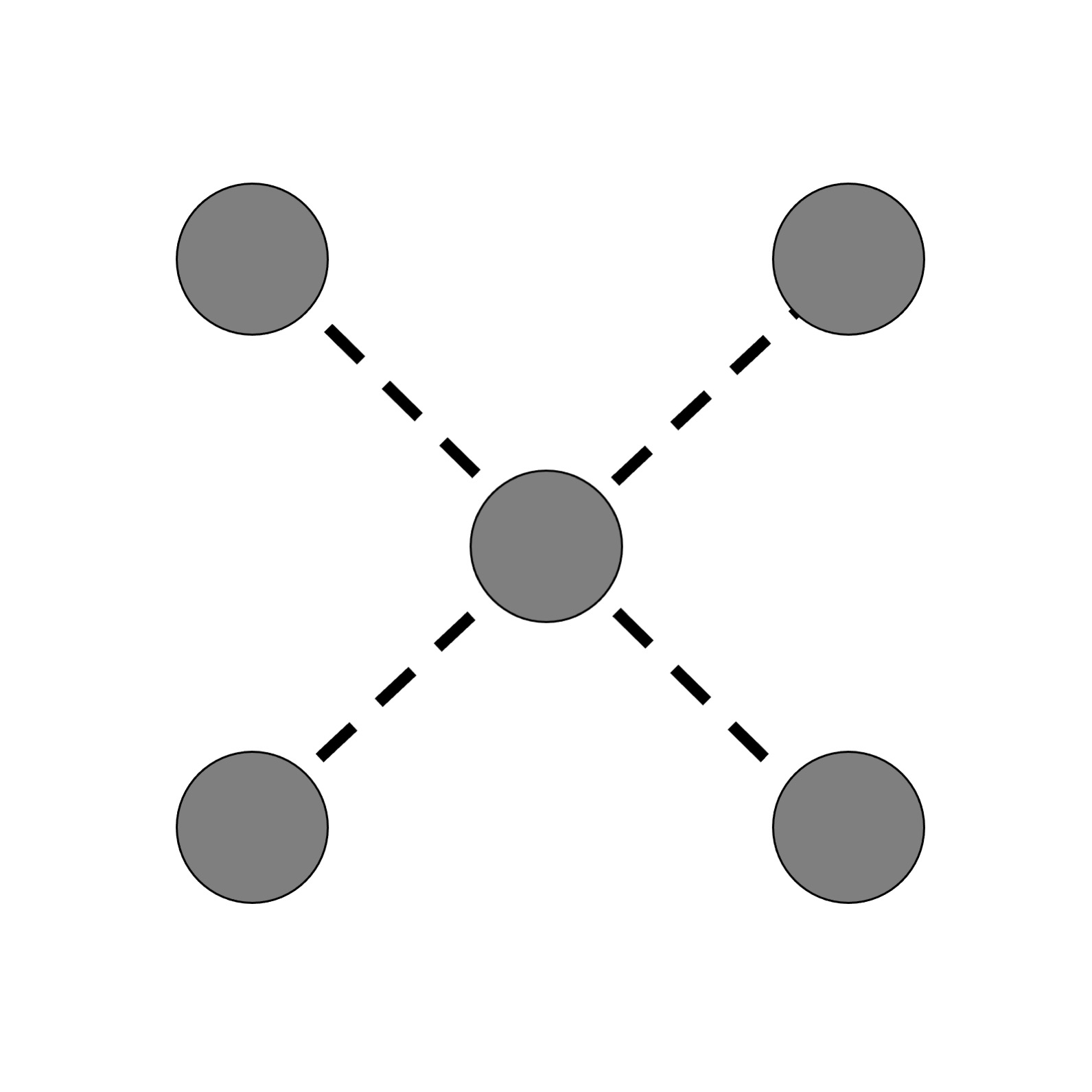} \\
\end{tabular}
\caption{The three connected 5-body shapes considered in this work.}
\label{fig:5b_systems}
\end{figure}

The specific 5-body configurations we study are the 5-full,
5-circle, and 5-star graphs as shown in~\figref{fig:5b_systems}.
The reasons for this selection are
\begin{itemize}
\item[(i)]
  to verify that the 5-full and 5-star configurations are renormalized
  by the $d_{\Delta} (\lambda)$ and $d_{\Lambda}(\lambda)$ couplings,
  respectively, and
\item[(ii)] to further confirm
  the exceptional status of the circle in the 5-body case, \ie,
  the fact that its proper renormalization requires
  the $d_{\Delta} (\lambda)$ coupling despite not
  containing any $\Delta$ subgraph.
\end{itemize}
The two hypotheses do indeed hold: in the case of the 5-full and 5-circle
configurations we numerically find the binding energies to be
\begin{eqnarray}
  E_{{\rm 5-full}} \approx 10.8\,E_{3 \Delta}\, , \\
  E_{{\rm 5-circle}} \approx 0.06\,E_{3 \Delta}\, .
\end{eqnarray}
We note, again, that the 5-circle cannot decay into other bound states.
Hence, there is no problem with its binding energy being
smaller than the 3-body $\Delta$ system.
For the 5-star we have instead
\begin{eqnarray}
  E_{{\rm 5-star}} \approx 30\,E_{3 \Lambda}\, ,
\end{eqnarray}
which follows the trend of binding energies much larger than one, {\it cf.},
the 4-star and 4-line, Eqs.~(\ref{eq.4to3ratios2a}) and
(\ref{eq.4to3ratios2b}).

Both findings, those regarding the 4- and 5-circle, suggest
the following conjecture: few-body systems whose unitary interactions
form a graph containing a closed loop will be renormalized by the 3-body
force that renormalizes the 3-body $\Delta$-shaped system (which, incidentally,
can also be labeled as the 3-circle).
Intuitively, this observations should hold for unitary systems
containing 3-, 4-, or 5-circle subgraphs: if the circle
component is not renormalized properly, neither will be
the system of which it is a part of.
However, even though the general idea does not seem implausible,
we have not found a rigorous proof yet.

\section{Explaining the two 4-body renormalization patterns}
\label{sec:boundary}

The renormalization of the 4-body unitary systems falls
into two distinct patterns: $\Delta$- and $\Lambda$-like.
Here we present a heuristic argument of why this is the case.
We stress that this is not a rigorous derivation of those renormalization
patterns, and yet, the explanation we provide, though incomplete,
helps to clarify in which cases we should expect
each type of 3-body force.

For understanding the patterns we will consider a zero-range
2-body resonant interaction.
That is, we will be considering the $\lambda \to \infty$ (or zero-range) limit,
in which a 2-body resonant interaction is reduced to a boundary condition
of the wave function at the origin, that is:
\begin{eqnarray}\label{eq.betheBC}
  \frac{d}{d r_{l k}} \left[ r_{lk} \Psi_{N} \right] \Big|_{r_{lk} = 0} = 0 \, ,
\end{eqnarray}
for each $l$, $k$ particle pair for which the interaction is resonant
($\vec{r}_l$ and $\vec{r}_k$ are single-particle coordinates and
\mbox{$\vec{r}_{lk} = \vec{r}_l - \vec{r}_k$}).
The Faddeev-component expansion of the $N$-body wave function
\begin{eqnarray}
  \Psi_N = \sum_{ij} \psi_{ij}(\vec{r}_{ij}, \dots) \, ,
\end{eqnarray}
yields
\begin{eqnarray}
  \frac{d}{d r_{l k}} \left[ r_{lk} \psi_{lk} \right] +
  \sum_{ij \neq lk} \psi_{ij}(\vec{r}_{ij}, \dots) \Big|_{r_{lk} = 0} = 0 \, .
\end{eqnarray}
In general, this set of equations will simplify owing to symmetries
that reduce the number of independent Faddeev components.
For instance, in the $N$-boson system or in the full $N$-body unitary system,
all the Faddeev components will be identical, \ie, $\psi_{ij} = \psi$;
in the first case this happens because of Bose-Einstein symmetry and
in the second because of symmetry under the permutation group.

In the 3-body system it is well-known how to derive the Efimov scaling
from the boundary condition in Eq.~(\ref{eq.betheBC}).
Though we will not present here the full derivation
(which can be found in~\cite{PhysRevLett.71.4103}),
we will try to understand the general patterns leading to the different types of
discrete scaling in the 3-body system (\ie, the standard $22.7$ discrete
scaling for the 3-boson system and the $1986.1$ scaling for
the heteronuclear case with equal masses).
In the $\Delta$ case, if we particularize the boundary condition for $lk = 12$
as a reference, we get
\begin{eqnarray}
  &&
  \frac{d}{d r_{12}} \left[ r_{12} \psi (\vec{r}_{12}, \vec{\rho}_3) \right]
  \Big|_{r_{12} = 0}
  \nonumber \\ && \quad +
  \psi (\vec{r}_{23}, \vec{\rho}_1) \Big|_{r_{12} = 0} +
  \psi (\vec{r}_{31}, \vec{\rho}_2) \Big|_{r_{12} = 0}
  = 0 \, ,
  \label{eq:3-circle-efimov}
\end{eqnarray}
where we have already taken into account that the three Faddeev components
of the wave function are formally identical, and
$\vec{\rho}_{k} = \vec{r}_k - (\vec{r}_i + \vec{r}_j)/2$ (plus the condition
$i \neq j \neq k$) being one of the Jacobi coordinates.
After suitable manipulations, the previous boundary condition will lead to
the standard Efimov effect with a discrete scaling of $22.7$.
In the $\Lambda$ system, we obtain instead 
\begin{eqnarray}
  \frac{d}{d r_{12}} \left[ r_{12} \psi (\vec{r}_{12}, \vec{\rho}_3) \right]
  \Big|_{r_{12} = 0} +
  \psi (\vec{r}_{31}, \vec{\rho}_2) \Big|_{r_{12} = 0} = 0 \, , \nonumber \\
  \label{eq:3-dandelion-efimov}
\end{eqnarray}
which translates into a discrete scaling of $1986.1$.
The specific steps leading from Eqs.~(\ref{eq:3-circle-efimov}) and
(\ref{eq:3-dandelion-efimov}) to the respective discrete scale
symmetry of these two systems are well-known
(see, \eg,~Ref.~\cite{Naidon:2016dpf}).

This argument can be easily extended to the 4-body system.
We will illustrate the idea with the 4-full system for which
\begin{eqnarray}
  && \frac{d}{d r_{12}}
  \left[ r_{12} \psi (\vec{r}_{12}, \dots) \right] \Big|_{r_{12} = 0}
  \nonumber \\ && \quad + \sum_{ij \neq 12 }
  \psi (\vec{r}_{ij}, \dots) \Big|_{r_{12} = 0}
  = 0 \, .
\end{eqnarray}
Superficially, this boundary condition does not seem to be equivalent to
the corresponding ones in the 3-body case. A more careful
inspection reveals that this is not the case.
Indeed, the six Faddeev components can be subdivided further depending on
the different choices of Jacobi coordinates (\ie, into Faddeev-Yakubovsky
components).
To provide a concrete example, we might expand the $ij = 12$, $23$, and $13$
components as
\begin{eqnarray}
  \psi (\vec{r}_{12}, \dots) &=&
  \psi_K (\vec{r}_{12}, \vec{\rho}_3, \vec{\sigma}_4) +
  \psi_K (\vec{r}_{12}, \vec{\rho}_4, \vec{\sigma}_3) \nonumber \\ &+&
  \psi_H (\vec{r}_{12}, \vec{h}_{12-34}, \vec{r}_{34}) \, , \\
  \psi (\vec{r}_{23}, \dots) &=&
  \psi_K (\vec{r}_{23}, \vec{\rho}_1, \vec{\sigma}_4) +
  \psi_K (\vec{r}_{23}, \vec{\rho}_4, \vec{\sigma}_1) \nonumber \\ &+&
  \psi_H (\vec{r}_{23}, \vec{h}_{23-14}, \vec{r}_{14}) \, , \\
  \psi (\vec{r}_{31}, \dots) &=&
  \psi_K (\vec{r}_{31}, \vec{\rho}_2, \vec{\sigma}_4) +
  \psi_K (\vec{r}_{31}, \vec{\rho}_4, \vec{\sigma}_2) \nonumber \\ &+&
  \psi_H (\vec{r}_{31}, \vec{h}_{31-24}, \vec{r}_{24}) \, ,
\end{eqnarray}
where we use the customary subscripts $K$ and $H$
to indicate the K- and H-components.
For the $K$ set of Jacobi coordinates, we have $\vec{\rho}_{k(ij)} = \vec{r}_k - (\vec{r}_i + \vec{r}_j)/2$ (though usually the $ij$ subscript will be
dropped, as it always corresponds to that of the first coordinate
$\vec{r}_{ij}$) and $\vec{\sigma}_l = \vec{r}_l - (\vec{r}_i + \vec{r}_j + \vec{r}_k)/3$ with $i \neq j \neq k \neq l$, while for the $H$ set we only have
one new coordinate: $\vec{h}_{ij-kl} =  (\vec{r}_i + \vec{r}_j)/2 -  (\vec{r}_l + \vec{r}_l)/2$.

The bottom line is that the original $12$ boundary condition
can be subdivided into two different types
of
boundary conditions.
One type involves the K-components only: for instance, if we separate
the Faddeev-Yakubovsky components that contain the $\sigma_4$ Jacobi
coordinate, we get
\begin{eqnarray}
  && \frac{d}{d r_{12}} \left[ r_{12}\,
    \psi_K (\vec{r}_{12}, \vec{\rho}_3, \vec{\sigma}_4) \right]
  \Big|_{r_{12} = 0} \nonumber \\ && \quad +
  \psi_K (\vec{r}_{23}, \vec{\rho}_1, \vec{\sigma}_4) \Big|_{r_{12} = 0} +
  \psi_K (\vec{r}_{31}, \vec{\rho}_2, \vec{\sigma}_4) \Big|_{r_{12} = 0} = 0 \, .
  \nonumber \\
  \label{eq:4-circle-efimov}
\end{eqnarray}
If we ignore the $\vec{\sigma}_4$ coordinate (which does not play a direct role
in the boundary condition), one realizes that this type of boundary
condition is equivalent to the one in the $\Delta$ system
(Eq.~ (\ref{eq:3-circle-efimov})).
We thus expect it to generate the same type of Efimov scaling, namely $22.7$,
which also implies that the 3-body force that renormalizes the $\Delta$
would renormalize the complete system.
We also notice that any few-body configuration containing a $\Delta$ subsystem
should contain a version of the boundary condition above, implying the
$22.7$ discrete scaling factor and its eventual renormalizability
by means of the $d_{\Delta}(\lambda)$ 3-body force.
In addition, there will be a different boundary condition for the H-components,
which we do not write down here: this second boundary condition is not trivial
to interpret, but we suspect that it is related to the existence of
two different universal solutions for the 4-boson
system~\cite{Hammer:2006ct}.

The extension to the other 4-body systems that contain at least a $\Delta$
subsystem (\ie, a closed loop of three resonant interactions) is laborious
(as not all the Faddeev components are identical) but straightforward.
It eventually leads to the boundary condition of
Eq.~(\ref{eq:4-circle-efimov})~ from which we expect to obtain
the same discrete scaling as the $\Delta$ (\ie, $22.7$).
It is thus no surprise that these systems will be renormalized by the
$\Delta$ 3-body force.

The extension to the 4-star is trivial, the most important
difference being the reduced number of Faddeev components (4 instead of 6).
If we assume that only the $1j$ interactions with $j=2,3,4$ are resonant,
we have the boundary condition
\begin{eqnarray}
  && \frac{d}{d r_{12}} \left[ r_{12} \psi (\vec{r}_{12}, \dots) \right] \Big|_{r_{12} = 0}
  \nonumber \\ && \quad + \sum_{j \neq 1 }\psi (\vec{r}_{1j}, \dots) \Big|_{r_{12} = 0}
  = 0 \, .
\end{eqnarray}
After expanding the Faddeev components into Faddeev-Yakubovsky ones,
we will end up with the following boundary condition for the K-components
\begin{eqnarray}
  && \frac{d}{d r_{12}} \left[ r_{12}\,
    \psi_K (\vec{r}_{12}, \vec{\rho}_3, \vec{\sigma}_4) \right]
  \Big|_{r_{12} = 0} \nonumber \\ && \quad +
  \psi_K (\vec{r}_{31}, \vec{\rho}_2, \vec{\sigma}_4) \Big|_{r_{12} = 0} = 0 \, .
  \label{eq:4-dandelion-efimov}
\end{eqnarray}
This condition is formally identical to the one in the $\Lambda$-shaped
3-body system except for the additional coordinate $\vec{\sigma}_4$
(which, again, does not play a direct role in the boundary condition).
Thus the conclusion is that its discrete scaling behavior should be identical to
that of the $\Lambda$-shaped 3-body system from which it follows that
it is also renormalized by the $\Lambda$ 3-body force,
$d_{\Lambda}(\lambda)$.
Again, though laborious, this argument can be extended to any $N$-body system
for which the unitary graph is a tree.

Now the problem lies with $N$-body systems that contain a unitary loop
with more than three lines, \eg, the 4-circle.
The boundary condition for the 4-circle reads
\begin{eqnarray}
  && \frac{d}{d r_{12}} \left[ r_{12} \,
    \psi (\vec{r}_{12}, \dots) \right] \Big|_{r_{12} = 0}
  \nonumber \\ && \quad
  {+ \psi (\vec{r}_{23}, \dots) 
  + \psi (\vec{r}_{34}, \dots) 
  + \psi (\vec{r}_{41}, \dots)}\, \Big|_{r_{12} = 0}
  \nonumber \\ && \quad
  = 0 \, ,
\end{eqnarray}
which, if expanded and matched in terms of K-type Jacobi coordinates and
Faddeev-Yakubovsky components, will result in a variation of
the boundary condition of the 4-star,
\ie, in Eq.~(\ref{eq:4-dandelion-efimov}).
However, our concrete calculations with the 4-circle show otherwise.
This indicates that our argument is incomplete for the 4-circle,
which points toward some overlooked factors in the previous arguments.
We conjecture that the missing factor is most probably a non-optimal
choice of Jacobi coordinates and Faddeev-Yakubovsky components,
which ignores the cyclical permutation symmetry of this system.
Basically, what we have been able to show is that the Bethe-Peierls boundary
condition for the 4-circle contains a {\it boundary subcondition}
similar to that of the $\Lambda$-shaped 3-body system,
Eq.~(\ref{eq:3-dandelion-efimov}).
Yet, this does not preclude the possibility that a different choice of
coordinates and components might uncover the presence of the more
stringent boundary condition of the $\Delta$-shaped 3-body system.
Indeed we know this to be the case owing to our numerical finding that
the 4-circle requires the $d_{\Delta}(\lambda)$ 3-body force
for its proper renormalization.
This situation also repeats itself for the 5-circle, which
leads us to the conjecture that all unitary clusters containing
a closed unitary interaction loop should have the same scaling
as in the standard Efimov effect, \ie, $22.7$.

Nonetheless, some unclear signals come from previous works in this sense.
In ref.~\cite{Naidon:2018}, the 4-circle has been studied using the same
renormalization scheme as for the lambda system.
The energy of the 4-circle was found to be unnaturally large, $\sim$400 E$_3$, and thereby indicative for a collapsing system and an associated new scale.
However, in the mass-imbalanced 4-circle the Efimovian excitation structure was found compatible with the mass-imbalanced lambda-system discrete scaling, in contradiction with our conjecture.
This may be explained by a transition of the Efimovian structure from Delta to Lambda-like induced by an increasing mass ratio.
This explanation is consistent with the more recent ref.~\cite{Frederico:2019bnm} where it was found that the geometric scaling
of the 4-body two-species system with a mass imbalance is smaller than the corresponding scaling of its 3-body counterpart. 
However, the arguments in~\cite{Frederico:2019bnm} depend explicitly
on the existence of a large difference in masses between
the two species and are thus not directly applicable to
our 4-circle configuration.
We conclude that, despite there are no yet clear contradictions in the studies published up to now, more attention to the spectral structure of the 4-circle is needed.

\section{Conclusions}
\label{sec:conclusions}

We have analyzed partially unitary few-body systems in which all particles
have the same mass but not all interparticle interactions
are resonant, only a subset of them.
Each of these few-body systems can be characterized
by a unitary graph, whose vertices and lines
represent particles and resonant interactions, respectively.
The resonant interactions 
can be modeled, without loss of generality,
by zero-range potentials, 
which are singular and
require regularization and renormalization.
Here, we analyzed the renormalization of the ground states of systems with $N=3$ and $4$ particles
(plus a few illustrative $N=5$ configurations).

From a series of numerical calculations and qualitative arguments,
we conjecture a relation between the geometry of the unitary graph
representing the partially unitary system
and its renormalization.
Partially unitary few-body systems do display Thomas collapse, \ie,
the binding energies of these systems diverge as the range of
their interactions approach zero.
As in the 3-body case, this collapse is avoided by the inclusion of a
zero-range, repulsive 3-body force which stabilizes
the binding energy of the ground state.
The type of 3-body force renormalizing a partially unitary 4-body
system (of which there are six, see \figref{tab.4topos}) depends
on the properties of the unitary graph of the latter:
\begin{itemize}
\item[(i)] 
  unitary tree-like graphs require the 3-body force
  that renormalizes the 3-body system with two resonant pairs
  (which we have called the $\Lambda$ system),
\item[(ii)] 
  unitary graphs containing closed loops are renormalized
  instead by the 3-body force of the 3-body system with
  three resonant pairs (which we have called the $\Delta$ system).
\end{itemize}
We have deduced (and verified) this renormalization pattern numerically
for each of the 4-body unitary graphs.
We conjecture that this pattern extends to partially unitary systems
with $N > 4$, though we have only verified this generalization
numerically for three selected $N=5$ systems.

Furthermore,
we have proposed a heuristic argument that exploits the representation
of a resonant pair in terms of a Bethe-Peierls boundary condition to show
that the different 4-body unitary graphs are indeed expected to be
renormalized by the 3-body force of the $\Lambda$- or
$\Delta$-shaped systems.
More specifically, what we have shown is that the resonant 2-body interaction
imposes a constraint on the 4-body wave function that is identical to
the analogous constraint for the 3-body system (modulo the presence
of an additional coordinate for the extra particle).
Incidentally, this is the same constraint that generates the characteristic
discrete scale invariance of $22.7$ and $1986.1$ for the $\Delta$- and
$\Lambda$-shaped 3-body systems.
Hence, we conjecture that the discrete scaling properties of
a particular unitary graph will follow one of these two patterns
(depending on whether they are renormalized by the 3-body force of
the $\Delta$- or $\Lambda$-shaped systems).
However, the argument fails for the particular case of cyclic graphs
-- the $N$-circles in the naming convention we follow -- which
indicates that our explanation of this behavior is incomplete,
representing an intriguing open problem
which is left to future work.

\section*{acknowledgement}
We owe thanks to H.~W.~Grie\ss hammer and N.~Barnea for insightful
discussions and critical comments.
J.K. was supported by the STFC~ST/P004423/1, the US Department of Energy
under contract DE-SC0015393, and the
hospitality of The George Washington University
and The Hebrew University.
M.P.V is partly supported by the National Natural Science Foundation
of China under Grants No. 11735003, No. 11835015, No. 11975041, No. 12047503
and No. 12125507, the Fundamental Research Funds for the Central
Universities and the Thousand Talents Plan
for Young Professionals.
M.P.V. would also like to thank the IJCLab of Orsay, where part of
this work has been done, for its long-term hospitality.

\bibliographystyle{apsrev}
\bibliography{SOE.bib}
\end{document}